\documentclass{aa}

\usepackage{txfonts}
\usepackage{natbib}
\bibpunct{(}{)}{;}{a}{}{,}
\usepackage[colorlinks=true,linkcolor=blue,citecolor=blue]{hyperref}
\usepackage{color}

\begin{document}



  \title{Formation of Wind-Fed Black Hole High-mass X-ray Binaries: \\ The Role of Roche-lobe-Overflow Post Black-Hole Formation}

  \author{Zepei\,Xing
          \inst{1,2}\fnmsep\thanks{e-mail: Zepei.Xing@unige.ch},
          Tassos\,Fragos
          \inst{1,2},
          Emmanouil\,Zapartas
          \inst{3},
          Tom\,M.\,Kwan
          \inst{4},
          Lixin\,Dai
          \inst{4},
          Ilya\,Mandel
          \inst{5,6},
          Matthias\,U.\,Kruckow
          \inst{1,2},
          Max\,Briel
          \inst{1,2},
          Jeff\, J.\,Andrews
          \inst{7},
          Simone\,S.\,Bavera
          \inst{1,2},
          Seth\,Gossage
          \inst{8},
          Konstantinos\,Kovlakas
          \inst{11,12},
          Kyle\,A.\,Rocha
          \inst{8,9},
          Meng\,Sun
          \inst{8},
          Philipp\,M.\,Srivastava
          \inst{8,10},
          }
  \authorrunning{Xing et al.}
    \titlerunning{Formation of Wind-Fed BH-HMXBs}
  \institute{
    Département d’Astronomie, Université de Genève, Chemin Pegasi 51, CH-1290 Versoix, Switzerland
   \and
   Gravitational Wave Science Center (GWSC), Université de Genève, CH1211 Geneva, Switzerland
   \and
   Institute of Astrophysics, FORTH, N. Plastira 100,  Heraklion, 70013, Greece
   \and
   Department of Physics, The University of Hong Kong, Pokfulam Road, Hong Kong
   \and
   School of Physics and Astronomy, Monash University, VIC 3800, Australia
   \and
   OzGrav: The Australian Research Council Centre of Excellence for Gravitational Wave Discovery, Clayton, VIC 3800, Australia
   \and
   Department of Physics, University of Florida, 2001 Museum Rd, Gainesville, FL 32611, USA
   \and
    Center for Interdisciplinary Exploration and Research in Astrophysics (CIERA), Northwestern University, 1800 Sherman Ave, Evanston, IL 60201, USA
   \and
   Department of Physics \& Astronomy, Northwestern University, 2145 Sheridan Road, Evanston, IL 60208, USA
   \and 
   Electrical and Computer Engineering, Northwestern University, 2145 Sheridan Road, Evanston, IL 60208, USA
   \and
   Institute of Space Sciences (ICE, CSIC), Campus UAB, Carrer de Magrans, 08193 Barcelona, Spain
   \and
   Institut d’Estudis Espacials de Catalunya (IEEC), Carrer Gran Capit\`a, 08034 Barcelona, Spain
   }
   \date{Received XXYYZZ; Accepted XXYYZZ}




\abstract{The three dynamically confirmed wind-fed black hole high-mass X-ray binaries (BH-HMXBs) are suggested to all contain a highly spinning black hole (BH). However, based on the theories of efficient angular momentum transport inside the stars, we expect that the first-born BHs in binary systems should have low spins, which is consistent with gravitational-wave observations. As a result, the origin of the high BH spins measured in wind-fed BH-HMXBs remains a mystery. In this paper, we conduct a binary population synthesis study on wind-fed BH-HMXBs at solar metallicity with the use of the newly developed code \texttt{POSYDON}, considering three scenarios for BH accretion: Eddington-limited, moderately super-Eddington, and fully conservative accretion. Taking into account the conditions for accretion-disk formation, we find that regardless of the accretion model, these systems are more likely to have already experienced a phase of Roche-lobe overflow after the BH formation. To account for the extreme BH spins, highly conservative accretion onto BHs is required, when assuming the accreted material carries the specific angular momentum at the innermost stable orbit. Besides, in our simulations we found that the systems with donor stars within the mass range of $10-20\,M_{\odot}$ are prevalent, posing a challenge in explaining simultaneously all observed properties of the BH-HMXB in our Galaxy, Cygnus X-1, and potentially hinting that the accretion efficiency onto non-degenerate stars, before the formation of the BH, is also more conservative than assumed in our simulations.}

\maketitle

\section{Introduction} \label{sec:intro}

Black hole X-ray binaries contain a black hole (BH) accreting material from a companion star, generating X-ray emission. Depending on the donor star mass and accretion processes, black hole X-ray binaries can be classified as low-mass X-ray binaries (BH-LMXBs) and high-mass X-ray binaries (BH-HMXBs). BH-LMXBs consist of a low-mass donor star ($\sim 2\,M_{\odot}$) that transfers mass through Roche lobe overflow onto the BH. BH-HMXBs typically involves a massive ($\gtrsim 10\,M_{\odot}$) OB-type donor star, with the BH accreting a fraction of the companion's stellar wind. Only three wind-fed BH-HMXBs have been dynamically confirmed: Cygnus X-1, LMC X-1 and M33 X-7 (we list their properties in Table \ref{HMXB}). All of them are observed to contain a BH with a high spin and a donor star that is still on its main sequence (MS), almost filling its Roche lobe. Besides, there are several candidates including SS433 \citep{2010ApJ...722..586S}, Cygnus X-3 \citep{2013MNRAS.429L.104Z}, HD96670 \citep{2021ApJ...913...48G}, IC10 X-1 \citep{2007ApJ...669L..21P,2015MNRAS.452L..31L}, and NGC300 X-1 \citep{2010MNRAS.403L..41C}, whose nature remains debated.

To explain why BH-HMXBs are much less numerous than Wolf-Rayet $+$ O star binaries in our Galaxy --- the latter being thought to be the progenitor of the former --- it has been proposed that the binaries are detectable as BH-HMXBs only when the stellar winds carry sufficient angular momentum to form an accretion disk \citep[e.g.][]{2020A&A...636A..99V,2021A&A...652A.138S}. More recently, \citet{hirai_mandel_2021}, through hydro-dynamical simulations, found that an accretion disk is formed when the MS donor's Roche lobe filling factor, $f_{\rm RL}$, defined as the ratio between the stellar radius and its Roche lobe radius, is $\gtrsim 0.8-0.9$, providing an explanation for the high $f_{\rm RL}$ of all three known systems. \citet{2024arXiv240608596S} analytically derived a disk formation criterion that does not constrain $f_{\rm RL}$. They predicted a subgroup of X-ray bright wind-fed BH-HMXBs in relatively wide orbits, containing B-type donor stars with low $f_{\rm RL}$. The disk can form due to the low-velocity winds of these stars. However, no such systems have been observed, and the overabundance of bright BH-HMXBs in their simulations might be related to the disk formation criterion. 

Most intriguingly, based on either the disk continuum fitting or the reflection line fitting methods, the BHs in these systems are almost maximally spinning \citep[see, e.g.][and references therein]{2014SSRv..183..295M,2014SSRv..183..277R,2021ARA&A..59..117R}. However, these methods are subject to potential uncertainties, which will be discussed in section ~\ref{sec:discussion_spin}.
The binary black hole (BBH) population from gravitational-wave observations indicates low aligned spins for BHs \citep[e.g.,][]{2019ApJ...882L..24A,2023PhRvX..13a1048A,2020ApJ...895..128M,2021PhRvD.104h3010R}. Arguments have been made that BH-HMXBs and BBHs originate from two distinct binary populations \citep{2022ApJ...929L..26F} and due to selection effects, it is natural that these two populations are showing different properties \citep{2023ApJ...946....4L}. Binary population synthesis (BPS) studies have shown that only a small fraction of BH-HMXBs result in BBH mergers \citep{2021ApJ...908..118N,2022ApJ...938L..19G, 2023MNRAS.524..245R}. 

However, the question of where the extreme spins come from has not yet been answered. The prescriptions of efficient angular momentum (AM) transport within stellar interiors, such as Tayler-Spruit (TS) dynamo \citep{2002A&A...381..923S}, the revised prescription by \citet{2019MNRAS.485.3661F}, and the calibrated TS dynamo presented in \citet{2022A&A...664L..16E}, can explain the rotation profile of the Sun \citep{2005A&A...440L...9E} and the core-rotation rate of low-mass red giants and subgiants from asteroseismic measurements \citep{2012Natur.481...55B,2012A&A...548A..10M,2014A&A...564A..27D,2018A&A...616A..24G}. These theories predict low natal BH spins in general \citep{2019ApJ...881L...1F}, but especially so for the first-born BHs in binary systems \citep{2015ApJ...800...17F}. The second-born BHs can obtain spins through tidal interactions between their progenitors and the first-born BHs, which usually occur in tight orbits following common-envelope events. In the framework of efficient AM transport, the effective inspiral spin distribution of BBHs can be reproduced in the isolated binary evolution channel \citep{2018A&A...616A..28Q,2020A&A...635A..97B,2021A&A...647A.153B,2020A&A...636A.104B,2021ApJ...921L...2O}. 

Looking at BH-HMXBs, \citet{2019ApJ...870L..18Q} investigated two possible scenarios that can produce spinning BHs, the case~A mass transfer (MT) channel and the chemical homogeneous evolution channel, and concluded that only if the internal AM transport is inefficient during the late evolution of the progenitor stars can high-spin BHs be produced, a scenario not supported yet by any other observational evidence or theoretical arguments. 
Another way for BHs to acquire significant spins is through super-Eddington accretion \citep{2011MNRAS.413..183M}. 
The observed, isotropic-equivalent, X-ray luminosity of ultraluminous X-ray sources \citep[e.g.][]{2006ApJ...649..730W,2020MNRAS.498.4790K} can exceed the Eddington limit by a few orders of magnitude. However, the possible geometrical beaming of the X-ray emission means that the actual accretion rate onto the BH does not have to be significantly super-Eddington \citep{2001ApJ...552L.109K,2007MNRAS.377.1187P,2022A&A...665A..28K}. \citet{2022RAA....22c5023Q} explored the possibility of highly super-Eddington accretion and its impact on BH spin in the context of BH-HMXBs. This study, however, was limited in that they only considered one specific binary configuration, did not model the evolution of the binary prior to the BH formation and neglected the fact that observed BH-HMXBs, like Cygnus X-1, are currently detached.


In this work, we use the newly-developed BPS code \texttt{POSYDON} \citep{2023ApJS..264...45F}, which incorporates extensive grids of detailed stellar structure and binary evolution models, to investigate the population of wind-fed BH-HMXBs.
Among other advantages, detailed binary evolution simulations can accurately model the case~A mass transfer phase. The case~A mass transfer phase after the BH formation is of great importance for the formation of wind-fed BH-HMXBs, but often treated poorly in rapid BPS codes due to the lack of information about the internal structure of stars \citep{2024MNRAS.tmp..142D}. 

The paper is organized as follows. In section \ref{sec:method} we briefly introduce the new binary-star model grids and the details of our BPS study. In section \ref{sec:result} we present the results of our binary models and population properties. We discuss the implications of this work on wind-fed BH-HMXBs in section \ref{sec:discussion} and summarize the conclusions of this paper in section \ref{sec:conclusion}. 

\begin{table*}[]
\centering
\renewcommand{\arraystretch}{1.5}
\caption{Properties of Wind-fed BH-HMXBs\label{HMXB}}
\begin{tabular}{l|ccccccc}
\hline\hline
BH-HMXB& $M_{\rm BH}\,[M_{\odot}]$ & $M_{\rm donor}\,[M_{\odot}]$ &  $P_{\rm{orb}}\,[\rm d]$ & $\chi_{\rm BH}$ &$f_{\rm RL}$ &$L_{\rm x}\,[\rm erg\,s^{-1}]$ & Reference\\
\hline

Cygnus X-1 & $21.2\pm{2.2}$ & $40.6^{+7.6}_{-7.1}$  & $5.60$ & $0.95^{+0.04}_{-0.084}$ & 0.997 & $2.1\times10^{37}$ & 1, 2, 3\\ 
LMC X-1 & $10.91\pm{1.41}$ & $31.79\pm{3.48}$  & $3.91$& $0.897^{+0.077}_{-0.176}$ &0.971 & $2.3\times10^{38}$  &3, 4\\
M33 X-7 & $11.4^{+3.3}_{-1.7}$ &$38^{+22}_{-10}$ & $3.45$& $0.84\pm{0.05}$\tablefootmark{a} &0.899& $(0.5-2)\times10^{38}$  &5, 6, 7\\
\hline
\end{tabular}

\tablefoot{\tablefoottext{a}{\citet{2022} updated a BH spin of $\approx 0.6$}}
\tablebib{(1)\citet{2011ApJ...742...84O}; (2)\citet{2021Sci...371.1046M}; (3)\citet{2023arXiv231116225D};
(4)\citet{2009ApJ...697..573O}; (5)\citet{2006ApJ...646..420P}; (6)\citet{2008ApJ...679L..37L,2010ApJ...719L.109L}; (7)\citet{2022}}

\end{table*}

\section{Method} \label{sec:method}

For all the computations presented in this paper, we use the publicly available BPS code \texttt{POSYDON}\footnote{We utilized the version of the  {\tt POSYDON} code identified by the commit hash 83a458d9d7c14cea2e371a8857bd061ac8c555c0, available at \href{https://github.com/POSYDON-code/POSYDON/}{https://github.com/POSYDON-code/POSYDON/}, along with the {\tt POSYDON} v1.0 dataset published at \href{https://zenodo.org/records/6655751}{https://zenodo.org/records/6655751}.}, which incorporates single- and binary-star model grids at solar metallicity simulated with the stellar evolution code Modules for Experiments in Stellar Astrophysics \citep[MESA,][]{2011ApJS..192....3P,2013ApJS..208....4P,2015ApJS..220...15P,2018ApJS..234...34P,2019ApJS..243...10P, 2023ApJS..265...15J}. The specific details of the stellar and binary physics employed in the production of precalculated grids are outlined in \citet{2023ApJS..264...45F}. Here we highlight the wind prescription adopted in our models. We utilized the Dutch scheme in \texttt{MESA} for stars with initial mass exceeding $8\,M{\odot}$. Specifically, we applied the wind-loss prescription of \citet{1988A&AS...72..259D} for cool stars with effective temperatures below $10,000\,\rm{K}$, and \citet{2001A&A...369..574V} scheme for hot stars with effective temperatures above $11,000\,\rm{K}$. For temperatures between these two thresholds, we implemented a linear interpolation of the wind-loss rates. Additionally, for hot stars with a surface hydrogen fraction below $0.4$, we adopted the Wolf-Rayet wind prescription proposed by \citet{2000A&A...360..227N}.

By default, in \texttt{POSYDON}, accretion onto BHs is assumed to be Eddington-limited. As part of this work, beyond the publicly available grids, we compute two additional binary-star model grids of binaries consisting of a compact object and a hydrogen rich, MS star (CO-HMS grids), where we vary the accretion efficiency onto BHs. The two additional grids share the same initial parameters and input physics with the original grid.
In the first new grid, we consider the limiting case of fully conservative accretion onto the BH, while in the second one, the accretion efficiency $\eta$ is established based on general relativistic radiation magnetohydrodynamic (GRRMHD) simulations \citep{Kwan.prep}. In that study, the authors carried out a series of simulations of super-Eddington disks in the magnetically arrested disk (MAD) state following the set up in \citet{Dai18} and \citet{Thomsen22} but around stellar-mass BHs. They varied the BH mass, spin and accretion rate $\dot{M}_{\rm acc}$ achieved in equilibrium states across different simulations to explore how these parameters affect the energy output and outflow properties. Based on the simulation results, they obtained an averaged fit for the relation between $\eta$ and the MT rate $\dot{M}_{\mathrm{tr}}$ (which is assumed to equal the sum of the mass accretion rate and the outflow rate obtained in the GRRMHD simulation):

\begin{equation}
    \eta \left(\dot{m}\right) \equiv \frac{\dot{M}_{\mathrm{acc}}}{\dot{M}_{\mathrm{tr}}} = 0.33 \left(\dot{m}\right)^{-0.08},
\end{equation}
for $\dot{m}\geq 10$, where $\dot{m}$ is the MT rate in units of the Eddington accretion rate $\dot{M}_{\mathrm{Edd}}$. In cases where $0< \dot{m} < 10$, we adopted an accretion rate of $\mathrm{min}(\dot{M}_{\mathrm{tr}},10\,\dot{M}_{\mathrm{Edd}}\,\eta(10))$. Generally, $\eta$ ranges within $10-30\%$ and is inversely correlated with $\dot{M}_{\mathrm{tr}}$.

For our synthetic HMXB population models, we simulate $500,000$ binaries for each accretion model, starting from two stars at zero-age MS. The primary star masses range within $10-120\,M_{\odot}$, with its distribution determined by the initial mass function from \citet{2001MNRAS.322..231K}. The secondary masses fall within the interval of $7-120\,M_{\odot}$, following a flat distribution from the minimum to the primary star mass. The initial orbital period distribution is uniform in logarithmic space from $0.35\,\rm{d}$ to $6000\,\rm{d}$, which is an extended prescription from \citet{2013A&A...550A.107S}. We identify wind-fed BH-HMXBs by searching for detached binaries containing a BH and a companion star with its Roche lobe filling factor, $f_{\rm RL}$, falling between $0.9 \lesssim f_{\rm RL} < 1$, which is valid across most of the parameter variations explored by \citet{hirai_mandel_2021}. We set a minimal mass of $10\,M_{\odot}$ for the companion star during the wind-fed phase. In all presented population models, we used the nearest-neighbor scheme for the interpolation between models in a grid \citep[see Section~7 in][]{2023ApJS..264...45F}.  

To determine the explodability of the pre-supernova core and the BH mass from the core properties, we adopted the \citet{2020MNRAS.499.2803P} prescription and supernova engine from \citet{2016ApJ...821...38S}. We assume the supernova kick velocities of the BHs from core-collapse supernova follow a Maxwellian distribution with a velocity dispersion of $\sigma_{\mathrm{CCSN}}=265\,\mathrm{km\,s^{-1}}$ \citep{2005MNRAS.360..974H}, rescaled by a factor of $1.4\,M_{\odot}/M_{\mathrm{BH}}$. The natal spin of a BH is calculated based on the internal rotational profile of the BH progenitor star, right before core collapse \citep[see Section~8.3.4 in][]{2023ApJS..264...45F}. The further evolution of the BH spin parameter $\chi_{\rm BH}$ is calculated by assuming the accreted material carries the specific angular momentum at the innermost stable circular orbit (ISCO). After the accretion of an infinitesimal mass, the dimensionless BH spin parameter $\chi_{\rm BH,f}$ can be calculated as \citep{1970Natur.226...64B}:
\begin{equation}
    \chi_{\rm BH,f} = \frac{r_{\mathrm{ISCO,i}}^{1/2}}{3} \frac{M_{\mathrm i}}{M_{\mathrm f}}\biggl\{4-\left[3\, r_{\mathrm{ISCO,i}}\left(\frac{M_{\mathrm i}}{M_{\mathrm f}}\right)^2-2\right]^{1/2} \biggl\},
\end{equation}
for where $M_{\mathrm i}$ and $M_{\mathrm f}$ are the BH masses before and after accretion, respectively (with ($M_{\mathrm f}-M_{\mathrm i}) \ll\ M_{\mathrm f}$), $r_{\mathrm{ISCO,i}}$ is the initial normalized ISCO radius that can be calculated as $r_{\mathrm{ISCO,i}} = 3 + Z_{2} - [(3-Z_{1})(3+Z_{1}+2Z_{2})]^{1/2}$, with $Z_{1} = 1 + (1-\chi_{\rm{BH,i}}^{2})^{1/3}[(1+\chi_{\rm{BH,i}})^{1/3}+(1-\chi_{\rm{BH,i}})^{1/3}]$ and $Z_{2} = (3\chi_{\rm{BH,i}}^{2}+Z_{1}^{2})^{1/2}$. The BH spin is updated at every timestep for the binary-star models in the CO-HMS grid. 

To estimate the X-ray luminosity of wind-fed BH-HMXBs, we calculate the wind mass accretion rate $\dot{M}_{\rm{w,acc}}$ onto the BHs based on the results of \citet{hirai_mandel_2021}, who integrate the mass flux through the accretion radius in their simulations. The mass accretion fraction, defined as the ratio of the mass accretion rate to the total mass-loss rate from the donor star, exhibits a similar trend as the Bondi–Hoyle accretion \citep{1944MNRAS.104..273B}, albeit subjects to large model uncertainties related to wind acceleration parameters and Eddington factors \citep{hirai_mandel_2021}. As a rough estimate, we adopt the values of the mass accretion fraction at $f_{\rm RL}\to 1$ from \citet{hirai_mandel_2021} based on their reference parameter configuration, and subsequently interpolate or extrapolate across various mass ratios. Then, the X-ray luminosity is given as
\begin{equation}
    L_{\mathrm{x}} = \epsilon \dot{M}_{\rm{w,acc}} c^{2},
\end{equation}
where $\epsilon = 0.1$ is the radiative efficiency of accretion.

\section{Results} \label{sec:result}

\subsection{Individual Wind-fed BH-HMXBs \label{sec:results_individual}}
\begin{figure*}
\includegraphics[width=0.96\textwidth]{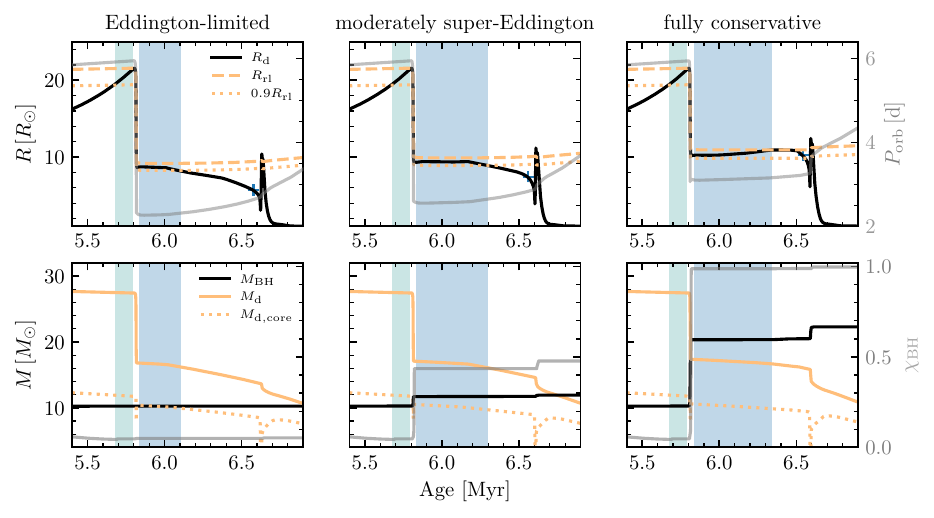}
\caption{Evolution of an example binary through the wind-fed HMXB phase under three BH accretion scenarios. The top panels show the evolution of the donor star radius (black solid line), Roche lobe radius (orange dashed line), 0.9 times the Roche lobe radius (orange dotted line), and orbital period (grey solid line). The green and blue areas indicate the duration of the wind-fed BH-HMXB phase before and after MT via RLO, respectively. The plus symbol denotes the end of the MS for the donor star. The bottom panels show the evolution of the BH mass (black solid line), donor star mass (orange solid line), donor star convective core mass (orange dotted line), and BH spin (grey solid line).
\label{fig:evolution}}
\end{figure*}

We first illustrate a reference binary-star evolutionary model, which at some point in its evolution appears as a bright wind-fed BH-HMXB system. The reference binary initially consists of a primary star of $34.9\,M_{\odot}$, a secondary star of $28.7\,M_{\odot}$, with an orbital period of $10.4\,\rm{d}$. The binary went through a case~A MT phase and subsequently a case~B MT phase. The primary star formed a BH of $12.2\,M_{\odot}$ with an initial spin of $0.047$. The secondary lost mass due to stellar winds but gained $\approx0.8\,M_{\odot}$ from accretion.
In Figure~\ref{fig:evolution}, we show the evolution of the binary from $\approx 0.3\,\rm{Myr}$ before it enters the wind-fed HMXB phase until the stage when the donor star undergoes significant contraction during its helium-burning phase. We can see that as the donor star evolves, it keeps expanding until it fills its Roche lobe. It stays at the wind-fed HMXB phase for a short period of $\approx 0.1\,\rm{Myr}$ before the Roche-lobe overflow (RLO) phase. After a fast RLO phase, the donor star loses nearly half of the envelope mass and its surface helium mass fraction increases by approximately a factor of two. This is consistent with the observations of Cygnus X-1, where the donor star's surface helium abundance is enhanced by more than a factor of two \citep{2012ARep...56..741S}. Its center helium mass fraction has increased to $0.9$, indicating that the star is approaching the end of its MS. At this stage, the change in the hydrogen and helium abundance profiles of the partially stripped star leads to changes in the mass-radius relation during the MS \citep[see, e.g.,][]{2019A&A...628A..19Q, 2022MNRAS.512.4116F}. The donor star contracts due to the mass loss and the binary becomes detached again. Then, the donor star stays very close to the Roche-lobe radius and the binary appears as a wind-fed HMXB for a longer period of $\approx 0.3-0.5\,\rm{Myr}$. Under the assumption of Eddington-limited accretion, the BH barely accretes any mass during the fast MT phase. Alternatively, the BH accretes $\simeq 2\,M_\odot$ and spins up to $\chi_{\rm BH} \approx 0.45$ assuming the moderately super-Eddington accretion scenario, while it reaches nearly maximum spin with fully conservative accretion. Because the transferred material that is not accreted by the BH leaves the system carrying away the BH's specific orbital AM, the Eddington-limited accretion leads to the largest binary orbital shrinkage due to MT. Moreover, Eddington-limited accretion results in the most significant stripping of the envelope, which subsequently leads to the quickest introduction of the Wolf-Rayet wind. Consequently, the duration of the second wind-fed phase is the shortest for the Eddington-limited accretion scenario. In the scenario of fully conservative accretion, the donor star experiences the least stripping and continues expanding after the fast RLO. As a result, following a long wind-fed phase, the binary enters a second slow RLO phase during the late MS stage of the donor star. Regardless of the differences, for all the three accretion scenarios, the binary spends more time as an observable wind-fed HMXB after the RLO phase, rather than before.

\subsection{Population of Wind-fed BH-HMXBs}

The individual binary models described in Section~\ref{sec:results_individual} provide us with some insights into the formation of wind-fed BH-HMXBs. However, it is a population study that can reveal whether the features identified in the evolution of a few individual binary models are representative of the whole population and allow us to evaluate their significance. Binaries that remain in a bright wind-fed BH-HMXB phase over a longer period of time are more likely to be detected. Consequently, we present distributions of the population properties for wind-fed BH-HMXBs in our simulations, weighted by the duration of the wind-fed HMXB phase for each binary. 

While in Section~\ref{sec:results_individual} we described the evolutionary history of one reference binary, several variations of evolutionary pathways are present in our BH-HMXB population models. For the moderately super-Eddington accretion model, for example, we found that, prior to BH formation, $48.3\%$ of binaries went through both case~A and case~B MT, $21.6\%$ only experienced case~A MT, $11.9\%$ avoided any MT, $11.6\%$ underwent case~B MT, and $6.5\%$ entered a contact phase. The first two channels dominate the population, characterized by similar initial conditions. The binaries that avoided MT contain a relatively massive primary star above $\approx 60\,M_{\odot}$ \citep{2023NatAs...7.1090B}. However, due to strong stellar winds, the BHs originating from these massive stars do not exhibit a significantly different mass distribution. Among these, the case~B MT channel is the only one showing a longer pre-RLO duration. These binaries tend to have wide orbits before RLO and more evolved secondary stars, resulting in a shorter post-RLO wind-fed phase.

\begin{figure*}[ht!]
\includegraphics[width=0.96\textwidth]{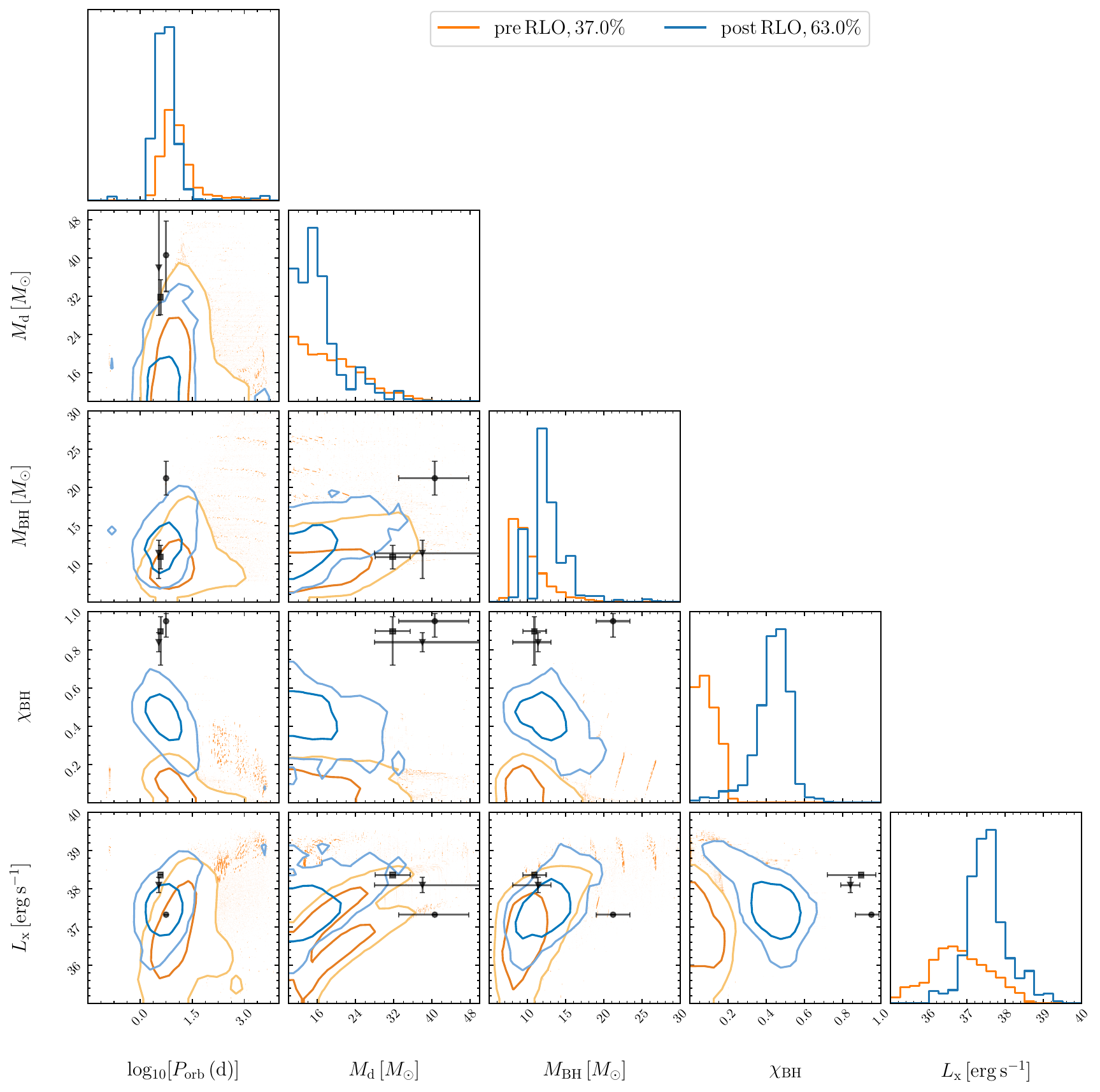}
\caption{Corner plot showing the time-weighted distributions of the properties of wind-fed BH-HMXBs from the moderately super-Eddington accretion scenario. The dark and light blue contours represent the $68\%$ and $95\%$ confidence regions, respectively, for the orbital period, donor star mass, BH mass, BH spin, and X-ray luminosity of systems after RLO. Similarly, the dark and light orange contours are for systems before RLO. The black dot, square, and triangle with error bars indicate the positions of Cygnus X-1, LMC X-1, and M33 X-7, respectively.
\label{fig:super}}
\end{figure*}

Figure~\ref{fig:super} shows the expected distributions of orbital periods, component masses, BH spins and X-ray luminosity for an observable wind-fed BH-HMXB population under the moderately super-Eddington accretion model. The figure distinguishes wind-fed BH-HMXBs that are observable before and after a RLO phase, shown with orange and blue color respectively. Notably, there is a higher probability of observing a wind-fed BH-HMXB after the occurrence of MT via RLO, constituting $63\%$ of the total duration.  \citet{2023MNRAS.524..245R} also mention that, in their BPS models, some binaries would engage in a second wind-fed BH-HMXB phase after MT via RLO, when the mass ratio is close to unity. Our detailed binary models show that, when weighted by duration, the binaries that are in the post-RLO wind-fed BH-HMXB phase are predominant, characterized by a broader range of mass ratios. This can be seen in the subplot of $M_{\rm{d}}$ versus $M_{\rm{BH}}$ of Figure~\ref{fig:super}, where the mass ratio can reach $\sim 2$. Compared to rapid BPS codes, which typically move across single stellar tracks during MT phases, \texttt{POSYDON} models the readjustment of mass-losing stars. Consequently, the partially stripped MS donor stars can shrink more substantially than rapid BPS codes predict in general, leading to a higher contribution from the post-RLO wind-fed BH-HMXB phase. The wind-fed BH-HMXB phase following RLO onto the BH constitutes an even higher fraction (almost 80\% by duration) of bright BH-HMXBs, with X-ray luminosities above $10^{37}\,\rm erg\,s^{-1}$ (see appendix~~\ref{sec:appendixc}).

As the MS donor stars should be in close orbits, nearly filling the Roche lobes, the majority of the binaries exhibits orbital periods of $\sim 1-20\,\rm{d}$, which is anticipated. The binaries that are in wide orbits extending to thousands of days in period contain post-MS donor stars. Those supergiant donor stars evolve much faster than MS stars, resulting in small contributions to the whole population. Besides, a smaller fraction of stripped stars, fitting in close orbits with orbital periods less than $1\,\rm{d}$, meet the requirement after experiencing common envelope evolution. Overall, MS donor stars account for $\approx 90\%$ of the whole population. However, it is important to note that \citet{hirai_mandel_2021} focused on line-driven winds and their conclusions about disk formation are only applicable to massive MS stars. Nevertheless, our population results indicate that post-MS donor stars are unlikely to make significant contributions due to their short evolutionary timescales, even if they have less strict requirements on the Roche-lobe filling factor for disk formation.

Regarding the donor star masses, there is a clear difference between the observations and the simulation samples. The three observed binaries contain donor stars above $\approx 30\,M_{\odot}$ while majority of donor stars are below $\approx 20\,M_{\odot}$ in our simulations. The three observed BH-HMXBs have different metallicities. As a result, direct comparison of donor star masses is not robust, as the simulations were conducted at solar metallicity. The implications of the distribution of the donor star masses are discussed in Section~\ref{sec:discussion}. The donor stars are mostly close to the terminal-age MS. We have included the distributions of the central and surface helium abundance of the donor stars in appendix~\ref{sec:appendixa}.

The BH masses are mostly within the range of $\sim 8-17\,M_{\odot}$, where moderately super-Eddington accretion has increased the current BH masses by a few solar masses compared to their birth values. For the BH spins, a bimodal distribution is evident, with the two peaks corresponding to the population before and after RLO. 
Finally, we show a rough estimate of the X-ray luminosity, acknowledging the model uncertainties associated with the accretion fraction calculated in \citet{hirai_mandel_2021}. We find that on average the post-RLO sub-population exhibits higher X-ray luminosities compared with the pre-RLO one. This is a result of the fact that on average the post-RLO sub-population has stronger stellar winds and higher mass ratios ($q=M_{\rm{BH}}/M_{\rm{d}}$) compared to the pre-RLO one.

\begin{figure*}[ht!]
\includegraphics[width=0.96\textwidth]{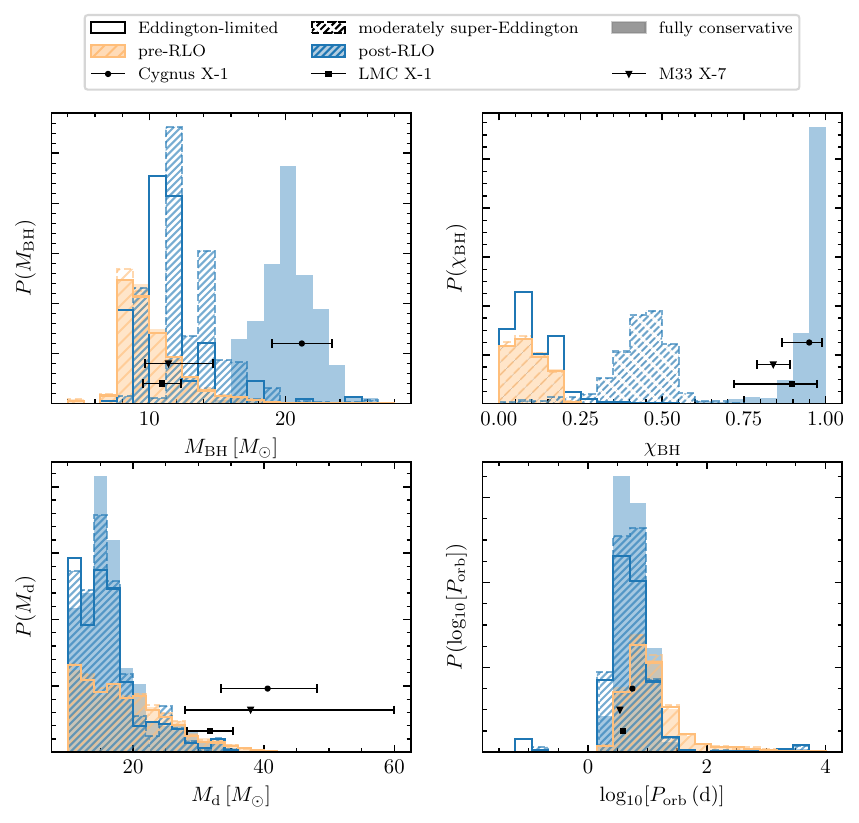}
\caption{Time-weighted probability distributions of the BH mass (top left), BH spin (top right), donor star mass (bottom left), and orbital period (bottom right) in wind-fed BH-HMXBs. The blue and orange indicate post-RLO and pre-RLO sub-populations, respectively. The unfilled bins, hashed bins, and filled bins represent the Eddington-limited, moderately super-Eddington, and fully conservative accretion scenarios, respectively. The black dot, square, and triangle with error bars, placed at arbitrary probability values, indicate the properties of Cygnus X-1, LMC X-1, and M33 X-7, respectively. 
\label{fig:spin}}
\end{figure*}

The HMXB population properties resulting from the other two accretion models can be found in appendix~\ref{sec:appendixb}. The most notable distinctions caused by accretion are evident in BH masses and spins. In Figure~\ref{fig:spin}, the top panels show the time-weighted probability distributions of the BH mass and spin for the three accretion scenarios, distinguishing between the wind-fed phases before and after MT via RLO. In the pre-RLO HMXB sub-populations, the distributions are the same, as expected. Most of the BH masses are below $\approx 15\,M_{\odot}$, and the spins are below $\approx 0.2$. In the case of Eddington-limited accretion, the BH mass and spin ranges remain nearly unchanged after RLO, but the mean of the BH mass shifts towards higher values. The higher normalization of the distribution denotes that on average the duration of the post-RLO wind-fed HMXB phase is longer, and that the probability of observing a wind-fed BH-HMXB after it has experienced a RLO phase is higher. Moderately super-Eddington accretion slightly increases the BH masses and spins up the BHs to a range of $\chi_{\rm{BH}} \sim 0.2-0.6$. For fully conservative accretion, a bimodal distribution of BH masses emerges, as the BHs could accrete a substantial amount of mass through RLO to be in a range of $\sim 15-25\,M_{\odot}$. After accretion, BH spins become extremely high, exceeding $\approx 0.9$ for the majority of them. 
In the bottom panels of Figure~\ref{fig:spin}, we present the probability distributions of the donor star mass and the orbital period, in which the three scenarios exhibit no significant differences. The fully conservative scenario leads to more stable MT, resulting in the disappearance of stripped stars coming from common envelope evolution. Regardless of the accretion scenarios, the total duration of the wind-fed BH-HMXB phase following MT via RLO is always longer than that before MT. The fraction of post-RLO duration relative to the total duration is greater than $60\%$ for all the three accretion scenarios.

\section{Discussion} \label{sec:discussion}
\subsection{Black Hole Spins}\label{sec:discussion_spin}
Our study has shown that, assuming efficient AM transport inside stars, the BHs in wind-fed BH-HMXBs are born with spins $\lesssim 0.2$. Eddington-limited accretion does not lead to significant changes in the spins. The moderately super-Eddington accretion model can produce BHs with spins of $\chi_{\rm{BH}} \sim 0.3-0.6$ whereas the fully conservative MT can spin up most of the BHs to extreme spins if the accreted material carries the specific AM at ISCO. It is evident that explaining the observed BH spins in BH-HMXBs poses a challenge in the absence of substantial accretion onto the BHs. Even when considering the impact of supernova processes on the natal BH spins, such as those arising from slow ejecta \citep{2018ApJ...862L...3S} or shock instability \citep{2016NewA...44...58M}, the attainment of extreme spins remains elusive. \citet{2017ApJ...846L..15B} found that BHs with high spins of $\sim 0.8$ can be produced, but only if the fallback material reaches the position of the companion star and extracts AM from the orbit.

The moderately super-Eddington accretion model is also insufficient to reproduce the extreme BH spins in BH-HMXBs. In the GRRMHD simulations, which are used to estimate the BH accretion efficiency in our moderately super-Eddington model, ordered magnetic fields thread the super-Eddington disks. These disks eventually enter MAD state, where the AM carried by the transferred material near the BH is less than the AM at the ISCO \citep[e.g.][]{Lowell.2024}. Therefore, the increase in BH spins in our simulations should be considered as an upper limit. Additionally, BHs fed by MADs launch powerful jets that can effectively spin down the BHs when the spin is very large \citep[][]{Tchekhovskoy.2012, Lowell.2024}. If the disks around the BHs in HMXBs easily reach the MAD state, it naturally excludes the possibility that the AM of the BH is obtained through gas accretion. However, if the disk does not contain enough magnetic flux to enter a MAD state, the accretion process can spin up the BH to a very high equilibrium spin value \citep{Gammie.2004}. Such a disk does not have significant magnetic pressure to power magnetic outflow, and it is expected that BHs accrete a higher fraction of the disk gas from non-MAD state super-Eddington disks, which lead to higher accretion efficiency and, thus, higher BH spins. In the fully-conservative accretion model, we find that a large fraction of BHs accreted more material than their initial masses, indicating that the accretion is more than adequate to produce extreme spins.

It is essential to note that the two primary methods for measuring BH spins in HMXBs may be susceptible to systematic uncertainties \citep{2018ApJ...855..120T,2021MNRAS.500.3640S,2021MNRAS.504.3424F} and model dependency \citep{2024ApJ...962..101Z}. For example, \citet{2024ApJ...967L...9Z} argue that when considering a warm, optically-thick, thermal Comptonizing layer on top of the disk, the spin of Cygnus X-1 is measured to be low, $\chi_{\rm{BH}} \lesssim 0.3$. Yet, if the extreme measured spins are real, to explain them in the context of super-Eddington accretion, mass and AM accretion efficiencies higher than those predicted by GRRMHD simulations of MAD-state super-Eddington disks should be considered.

If both the mass accretion and AM transport to BHs are efficient enough to produce highly spinning BHs, it would likely also influence the properties of coalescing BBH populations. During the evolution of merging BBHs, RLO accretion onto the first-born BH is anticipated. The BHs can undergo spin-up due to efficient accretion. In this case, we would not expect the first-born BHs to exhibit nearly-zero spins in BBH mergers. \citet{2022ApJ...930...26S} employed mildly super-Eddington accretion of case~A MT onto the BHs, highlighting its capability to account for the observed highly spinning BHs in BBH mergers. \citet{2021A&A...647A.153B} found that highly conservative accretion onto BHs effectively reduces the contribution from the stable MT channel for BBH mergers, as the binaries do not shrink sufficiently during MT to coalesce due to gravitational waves within the Hubble time. However, most rapid BPS codes, such as \texttt{COSMIC}, which was utilized by \citet{2021A&A...647A.153B} to model the MT phase, are recognized for their inadequate treatment of case~A mass transfer \citep{2024MNRAS.tmp..142D}. In contrast, BPS codes that utilize detailed binary models and assume fully conservative accretion onto black holes \citep[e.g.][]{2023MNRAS.520.5724B} indicate that the stable MT channel remains viable. However, a direct comparison between the models presented in these studies should be done with caution, as they also differ in several other key physics assumptions. 
The specific properties of BBHs, distribution of effective inspiral spins, and their formation channels under efficient BH accretion scenarios should be investigated in future studies.

\subsection{Donor Star Masses}

The three observed wind-fed BH-HMXBs exhibit donor stars with masses $\gtrsim 30\,M_{\odot}$. In our population studies, we found that donor stars within $\sim 10-20\,M_{\odot}$ are expected to be prevalent in wind-fed BH-HMXBs at solar metallicity. One caveat in comparing our population models to the observed properties of the three known wind-fed BH-HMXBs is that our population models assume solar metallicity, while the observed systems have varied metallicities. Specifically, M33 X-7 has been suggested to have a metallicity between $0.1\,Z_{\odot}$ \citep{2007Natur.449..872O} and $0.5\,Z_{\odot}$ \citep{2022}. Similarly, LMC X-1 is expected to have sub-solar metallicity, consistent with the LMC metallicity. For these two systems, comparisons with our simulations are not straightforward, as lower metallicities can lead to the formation and survival of more massive stars. For Cygnus X-1, however, a solar or even a supersolar metallicity \citep{2012ARep...56..741S} is suggested, making its $40\,M{\odot}$ donor star hard to reconcile with our simulations (but see discussion in \citet{2021ApJ...908..118N}).

The prevalence of $\sim 10-20\,M_{\odot}$ stars in our simulations can be explained as follows. Firstly, the accretion onto non-degenerate stars in \texttt{POSYDON} is regulated by stellar rotation, typically leading to low accretion efficiency \citep{2023ApJS..264...45F}. Consequently, the secondary stars generally do not gain substantial mass before the BH formation. Such inefficient accretion is suggested to prevent the formation of merging NSBH with first-born NSs \citep{2024A&A...683A.144X} and potentially contributes to discrepancies between the predicted and observed Be X-ray binary donor mass distributions \citep{2020MNRAS.498.4705V,2024arXiv240307172A}. In the case of wind-fed BH-HMXBs, if the mass accretion efficiency is higher during earlier evolutionary stages, the donor star could have accreted more mass before BH formation, potentially alleviating the tension between our simulations and the observations.

Secondly, at solar metallicity, when the primary star becomes a BH, the secondary star should have lost a significant fraction of its mass through stellar winds if it is initially a massive star $\gtrsim 40\,M_{\odot}$. Moreover, many BH-HMXBs should be in the second wind-fed phase after the donor stars have transferred mass to the BHs through RLO. \citet{2023MNRAS.524..245R} produced wind-fed BH-HMXBs with donor stars that can reach $\sim 70\,M_{\odot}$ at solar metallicity. In our simulations, when the BH forms, the massive donor star has lost a substantial amount of mass due to stellar winds and mass transfer, reducing its mass to below $\approx 50\,M_{\odot}$. The major difference arises from the modelling of stellar winds of massive stars at the MS stage. We found that massive stars $\gtrsim 80\,M_{\odot}$ at solar metallicity have strong winds that can deplete the hydrogen envelope and induce Wolf–Rayet winds during their MS stage \citep{2023NatAs...7.1090B}. As a result, the maximum donor star mass in our simulations is below $50\,M_{\odot}$. 

Moreover, according to the IMF, the number of low-mass stars surpasses the number of high-mass stars. With a flat distribution of the initial mass ratio and without any selection effect on the donor stars, it is natural to expect a prevalence of $\sim 10-20\,M_{\odot}$ donor stars as they are more common and have longer lifetimes compared with a $\sim 40\,M_{\odot}$ star. The absence of massive donor stars was also indicated by \citet{2024arXiv240608596S}, who found X-ray bright binaries do not include contributions from initially $\gtrsim 40\,M_{\odot}$ donor stars.

The differences between modelled and observed donor star masses might indicate that certain physical assumptions or treatments in our stellar and binary evolution are not robust. Wind prescriptions, mass accretion efficiency for non-degenerate stars, and AM loss from non-conservative MT, as suggested in \citet{2023MNRAS.524..245R}, could all impact the population properties. Therefore, it is imperative to investigate them further, seeking a consistent model that can be applied to diverse binary systems.

Another avenue worth exploring involves investigating the selection effects that favor massive donor stars. Our estimation of X-ray luminosity is too rough to be quantitatively analyzed as we only adopt one set of parameters for estimating the accretion efficiency for all binaries from \citet{hirai_mandel_2021} and interpolate or extrapolate among different mass ratios. Moreover, the mass accretion rate obtained in \citet{hirai_mandel_2021} is an upper limit because not all mass passing through the accretion radius is guaranteed to be accreted. In our simulations, most donor stars with masses within $\sim 10-20\,M_{\odot}$ can produce X-ray luminosity over $10^{37}\,\rm{erg\,s^{-1}}$. However, if we apply the low-end efficiency in \citet{hirai_mandel_2021} to all binaries, those with X-ray luminosity above $10^{37}\,\rm{erg\,s^{-1}}$ account for $\approx 21\%$ of the total population and $\approx 56\%$ of them contain a donor star greater than $20\,M_{\odot}$. A corner plot showing the property distributions for these X-ray bright binaries can be found in appendix~\ref{sec:appendixc}.

\section{Conclusions} \label{sec:conclusion}

The three observed wind-fed BH-HMXBs demonstrate high BH spins, the origin of which remains an open question. We used the new-generation BPS code \texttt{POSYDON} to investigate the population properties of wind-fed BH-HMXBs, adopting three CO-HMS grids with different assumed accretion efficiencies onto BHs. We summarize the key conclusions here:
\begin{enumerate}
\item Wind-fed BH-HMXBs at solar metallicity are more likely to have already been though a fast RLO phase, after the BH formation. Regardless of the mass-accretion models, in our simulations, the total duration of wind-fed BH-HMXBs in a post-RLO phase is higher than that in a pre-RLO phase. After mass transfer via RLO, the partially stripped donor star contracts due to the changes in its hydrogen and helium abundance profile, remaining close to its Roche lobe. At this stage, the binary stays detached for a long period, appearing as a wind-fed BH-HMXB. We highlight the importance of employing detailed modelling in BPS studies, which includes a self-consistent treatment of MT and subsequent readjustment of stars experiencing mass loss or gain.

\item An accretion efficiency prescription inspired by GRRMHD simulations can produce BHs with moderate spins in the range of $\sim 0.3-0.6$, while fully conservative MT can spin up BHs to near-maximal spins, when assuming that the accreted material carries the specific AM at the ISCO. To explain the high spins observed in the three wind-fed BH-HMXBs by super-Eddington accretion, a mass and AM accretion efficiency onto BHs higher that that predicted by GRRMHD simulation of super-Eddington disks in the MAD state is required.

\item In our simulations, MS donor stars with  masses in the range of $\sim 10-20\,M_{\odot}$ dominate the wind-fed BH-HMXB population at solar metallicity. This is in tension with the fact that the only confirmed wind-fed BH-HMXB in our Galaxy, Cygnus X-1, contains a donor star of $\approx40\,M_{\odot}$. Assuming the mass measurements are reliable, we need to explore further observational selection effects that favor massive donor stars, as well as consider an accretion efficiency model onto non-degenerate stars that leads to more conservative MT.
\end{enumerate}

\begin{acknowledgements}
The POSYDON project is supported primarily by two sources: the Swiss National Science Foundation (PI Fragos, project numbers PP00P2\_211006 and CRSII5\_213497) and the Gordon and Betty Moore Foundation (PI Kalogera, grant award GBMF8477). This project has received funding from the European Union’s Horizon 2020 research and innovation programme under the Marie Sklodowska-Curie RISE action, grant agreement  No 873089 (ASTROSTAT- II).
Z.X. acknowledges support from the China Scholarship Council (CSC).
E.Z. acknowledges funding support from the Hellenic Foundation for Research and Innovation (H.F.R.I.) under the "3rd Call for H.F.R.I. Research Projects to support Post-Doctoral Researchers" (Project No: 7933). 
S.S.B., T.F., and Z.X. were supported by the project number PP00P2\_211006. T.K. and L.D. acknowledge the support from the National Natural Science Foundation of China and the Hong Kong Research Grants Council (12122309, N\_HKU782/23, 17305523, 17314822).
S.S.B. was also supported by the project number CRSII5\_213497. 
M.B. ackowledges support from the Boninchi Foundation.
K.A.R.\ and M.S. are supported by the Gordon and Betty Moore Foundation (PI Kalogera, grant award GBMF8477)
K.A.R.\ is also supported by the Riedel Family Fellowship. 
K.A.R.\ also thanks the LSSTC Data Science Fellowship Program, which is funded by LSSTC, NSF Cybertraining Grant No.\ 1829740, the Brinson Foundation, and the Moore Foundation; their participation in the program has benefited this work.
K.K. acknowledges support from the Spanish State Research Agency, through the María de Maeztu Program for Centers and Units of Excellence in R\&D, No. CEX2020-001058-M. 
J.J.A.~acknowledges support for Program number (JWST-AR-04369.001-A) provided through a grant from the STScI under NASA contract NAS5-03127.  I.M.~acknowledges
support from the Australian Research Council Centre of
Excellence for Gravitational Wave Discovery (OzGrav),
through project number CE230100016.
\end{acknowledgements}

\bibliography{sample631}{}

\begin{thebibliography}{95}
\expandafter\ifx\csname natexlab\endcsname\relax\def\natexlab#1{#1}\fi

\bibitem[{{Abbott} {et~al.}(2019){Abbott}, {Abbott}, {Abbott}, {Abraham},
  {Acernese}, {Ackley}, {Adams}, {Adhikari}, {Adya}, {Affeldt}, {Agathos},
  {Agatsuma}, {Aggarwal}, {Aguiar}, {Aiello}, {Ain}, {Ajith}, {Allen},
  {Allocca}, {Aloy}, {Altin}, {Amato}, {Ananyeva}, {Anderson}, {Anderson},
  {Angelova}, {Antier}, {Appert}, {Arai}, {Araya}, {Areeda}, {Ar{\`e}ne},
  {Arnaud}, {Arun}, {Ascenzi}, {Ashton}, {Aston}, {Astone}, {Aubin}, {Aufmuth},
  {AultONeal}, {Austin}, {Avendano}, {Avila-Alvarez}, {Babak}, {Bacon},
  {Badaracco}, {Bader}, {Bae}, {Baker}, {Baldaccini}, {Ballardin}, {Ballmer},
  {Banagiri}, {Barayoga}, {Barclay}, {Barish}, {Barker}, {Barkett}, {Barnum},
  {Barone}, {Barr}, {Barsotti}, {Barsuglia}, {Barta}, {Bartlett}, {Bartos},
  {Bassiri}, {Basti}, {Bawaj}, {Bayley}, {Bazzan}, {B{\'e}csy}, {Bejger},
  {Belahcene}, {Bell}, {Beniwal}, {Berger}, {Bergmann}, {Bernuzzi}, {Bero},
  {Berry}, {Bersanetti}, {Bertolini}, {Betzwieser}, {Bhandare}, {Bidler},
  {Bilenko}, {Bilgili}, {Billingsley}, {Birch}, {Birney}, {Birnholtz},
  {Biscans}, {Biscoveanu}, {Bisht}, {Bitossi}, {Bizouard}, {Blackburn},
  {Blair}, {Blair}, {Blair}, {Bloemen}, {Bode}, {Boer}, {Boetzel}, {Bogaert},
  {Bondu}, {Bonilla}, {Bonnand}, {Booker}, {Boom}, {Booth}, {Bork}, {Boschi},
  {Bose}, {Bossie}, {Bossilkov}, {Bosveld}, {Bouffanais}, {Bozzi},
  {Bradaschia}, {Brady}, {Bramley}, {Branchesi}, {Brau}, {Briant}, {Briggs},
  {Brighenti}, {Brillet}, {Brinkmann}, {Brisson}, {Brockill}, {Brooks},
  {Brown}, {Brunett}, {Buikema}, {Bulik}, {Bulten}, {Buonanno}, {Buscicchio},
  {Buskulic}, {Buy}, {Byer}, {Cabero}, {Cadonati}, {Cagnoli}, {Cahillane},
  {Calder{\'o}n Bustillo}, {Callister}, {Calloni}, {Camp}, {Campbell},
  {Canepa}, {Cannon}, {Cao}, {Cao}, {Capocasa}, {Carbognani}, {Caride},
  {Carney}, {Carullo}, {Casanueva Diaz}, {Casentini}, {Caudill},
  {Cavagli{\`a}}, {Cavalier}, {Cavalieri}, {Cella}, {Cerd{\'a}-Dur{\'a}n},
  {Cerretani}, {Cesarini}, {Chaibi}, {Chakravarti}, {Chamberlin}, {Chan},
  {Chao}, {Charlton}, {Chase}, {Chassande-Mottin}, {Chatterjee}, {Chaturvedi},
  {Chatziioannou}, {Cheeseboro}, {Chen}, {Chen}, {Chen}, {Cheng}, {Cheong},
  {Chia}, {Chincarini}, {Chiummo}, {Cho}, {Cho}, {Cho}, {Christensen}, {Chu},
  {Chua}, {Chung}, {Chung}, {Ciani}, {Ciobanu}, {Ciolfi}, {Cipriano}, {Cirone},
  {Clara}, {Clark}, {Clearwater}, {Cleva}, {Cocchieri}, {Coccia}, {Cohadon},
  {Cohen}, {Colgan}, {Colleoni}, {Collette}, {Collins}, {Cominsky},
  {Constancio}, {Conti}, {Cooper}, {Corban}, {Corbitt}, {Cordero-Carri{\'o}n},
  {Corley}, {Cornish}, {Corsi}, {Cortese}, {Costa}, {Cotesta}, {Coughlin},
  {Coughlin}, {Coulon}, {Countryman}, {Couvares}, {Covas}, {Cowan}, {Coward},
  {Cowart}, {Coyne}, {Coyne}, {Creighton}, {Creighton}, {Cripe}, {Croquette},
  {Crowder}, {Cullen}, {Cumming}, {Cunningham}, {Cuoco}, {Dal Canton},
  {D{\'a}lya}, {Danilishin}, {D'Antonio}, {Danzmann}, {Dasgupta}, {Da Silva
  Costa}, {Datrier}, {Dattilo}, {Dave}, {Davier}, {Davis}, {Daw}, {DeBra},
  {Deenadayalan}, {Degallaix}, {De Laurentis}, {Del{\'e}glise}, {Del Pozzo},
  {DeMarchi}, {Demos}, {Dent}, {De Pietri}, {Derby}, {De Rosa}, {De Rossi},
  {DeSalvo}, {de Varona}, {Dhurandhar}, {D{\'\i}az}, {Dietrich}, {Di Fiore},
  {Di Giovanni}, {Di Girolamo}, {Di Lieto}, {Ding}, {Di Pace}, {Di Palma}, {Di
  Renzo}, {Dmitriev}, {Doctor}, {Donovan}, {Dooley}, {Doravari}, {Dorrington},
  {Downes}, {Drago}, {Driggers}, {Du}, {Ducoin}, {Dupej}, {Dwyer}, {Easter},
  {Edo}, {Edwards}, {Effler}, {Ehrens}, {Eichholz}, {Eikenberry}, {Eisenmann},
  {Eisenstein}, {Essick}, {Estelles}, {Estevez}, {Etienne}, {Etzel}, {Evans},
  {Evans}, {Fafone}, {Fair}, {Fairhurst}, {Fan}, {Farinon}, {Farr}, {Farr},
  {Fauchon-Jones}, {Favata}, {Fays}, {Fazio}, {Fee}, {Feicht}, {Fejer}, {Feng},
  {Fernandez-Galiana}, {Ferrante}, {Ferreira}, {Ferreira}, {Ferrini},
  {Fidecaro}, {Fiori}, {Fiorucci}, {Fishbach}, {Fisher}, {Fishner},
  {Fitz-Axen}, {Flaminio}, {Fletcher}, {Flynn}, {Fong}, {Font}, {Forsyth},
  {Fournier}, {Frasca}, {Frasconi}, {Frei}, {Freise}, {Frey}, {Frey},
  {Fritschel}, {Frolov}, {Fulda}, {Fyffe}, {Gabbard}, {Gadre}, {Gaebel},
  {Gair}, {Gammaitoni}, {Ganija}, {Gaonkar}, {Garcia},
  {Garc{\'\i}a-Quir{\'o}s}, {Garufi}, {Gateley}, {Gaudio}, {Gaur}, {Gayathri},
  {Gemme}, {Genin}, {Gennai}, {George}, {George}, {Gergely}, {Germain},
  {Ghonge}, {Ghosh}, {Ghosh}, {Ghosh}, {Giacomazzo}, {Giaime}, {Giardina},
  {Giazotto}, {Gill}, {Giordano}, {Glover}, {Godwin}, {Goetz}, {Goetz},
  {Goncharov}, {Gonz{\'a}lez}, {Gonzalez Castro}, {Gopakumar}, {Gorodetsky},
  {Gossan}, {Gosselin}, {Gouaty}, {Grado}, {Graef}, {Granata}, {Grant}, {Gras},
  {Grassia}, {Gray}, {Gray}, {Greco}, {Green}, {Green}, {Gretarsson}, {Groot},
  {Grote}, {Grunewald}, {Gruning}, {Guidi}, {Gulati}, {Guo}, {Gupta}, {Gupta},
  {Gustafson}, {Gustafson}, {Haegel}, {Halim}, {Hall}, {Hall}, {Hamilton},
  {Hammond}, {Haney}, {Hanke}, {Hanks}, {Hanna}, {Hannam}, {Hannuksela},
  {Hanson}, {Hardwick}, {Haris}, {Harms}, {Harry}, {Harry}, {Haster},
  {Haughian}, {Hayes}, {Healy}, {Heidmann}, {Heintze}, {Heitmann}, {Hello},
  {Hemming}, {Hendry}, {Heng}, {Hennig}, {Heptonstall}, {Hernandez Vivanco},
  {Heurs}, {Hild}, {Hinderer}, {Hoak}, {Hochheim}, {Hofman}, {Holgado},
  {Holland}, {Holt}, {Holz}, {Hopkins}, {Horst}, {Hough}, {Howell}, {Hoy},
  {Hreibi}, {Huerta}, {Huet}, {Hughey}, {Hulko}, {Husa}, {Huttner},
  {Huynh-Dinh}, {Idzkowski}, {Iess}, {Ingram}, {Inta}, {Intini}, {Irwin},
  {Isa}, {Isac}, {Isi}, {Iyer}, {Izumi}, {Jacqmin}, {Jadhav}, {Jani},
  {Janthalur}, {Jaranowski}, {Jenkins}, {Jiang}, {Johnson}, {Jones}, {Jones},
  {Jones}, {Jonker}, {Ju}, {Junker}, {Kalaghatgi}, {Kalogera}, {Kamai},
  {Kandhasamy}, {Kang}, {Kanner}, {Kapadia}, {Karki}, {Karvinen}, {Kashyap},
  {Kasprzack}, {Katsanevas}, {Katsavounidis}, {Katzman}, {Kaufer}, {Kawabe},
  {Keerthana}, {K{\'e}f{\'e}lian}, {Keitel}, {Kennedy}, {Key}, {Khalili},
  {Khan}, {Khan}, {Khan}, {Khan}, {Khazanov}, {Khursheed}, {Kijbunchoo}, {Kim},
  {Kim}, {Kim}, {Kim}, {Kim}, {Kim}, {Kimball}, {King}, {King},
  {Kinley-Hanlon}, {Kirchhoff}, {Kissel}, {Kleybolte}, {Klika}, {Klimenko},
  {Knowles}, {Koch}, {Koehlenbeck}, {Koekoek}, {Koley}, {Kondrashov}, {Kontos},
  {Koper}, {Korobko}, {Korth}, {Kowalska}, {Kozak}, {Kringel}, {Krishnendu},
  {Kr{\'o}lak}, {Kuehn}, {Kumar}, {Kumar}, {Kumar}, {Kumar}, {Kuo}, {Kutynia},
  {Kwang}, {Lackey}, {Lai}, {Lam}, {Landry}, {Lane}, {Lang}, {Lange}, {Lantz},
  {Lanza}, {Lartaux-Vollard}, {Lasky}, {Laxen}, {Lazzarini}, {Lazzaro},
  {Leaci}, {Leavey}, {Lecoeuche}, {Lee}, {Lee}, {Lee}, {Lee}, {Lee}, {Lee},
  {Lehmann}, {Lenon}, {Leroy}, {Letendre}, {Levin}, {Li}, {Li}, {Li}, {Li},
  {Lin}, {Linde}, {Linker}, {Littenberg}, {Liu}, {Liu}, {Lo}, {Lockerbie},
  {London}, {Longo}, {Lorenzini}, {Loriette}, {Lormand}, {Losurdo}, {Lough},
  {Lousto}, {Lovelace}, {Lower}, {L{\"u}ck}, {Lumaca}, {Lundgren}, {Lynch},
  {Ma}, {Macas}, {Macfoy}, {MacInnis}, {Macleod}, {Macquet},
  {Maga{\~n}a-Sandoval}, {Maga{\~n}a Zertuche}, {Magee}, {Majorana},
  {Maksimovic}, {Malik}, {Man}, {Mandic}, {Mangano}, {Mansell}, {Manske},
  {Mantovani}, {Mapelli}, {Marchesoni}, {Marion}, {M{\'a}rka}, {M{\'a}rka},
  {Markakis}, {Markosyan}, {Markowitz}, {Maros}, {Marquina}, {Marsat},
  {Martelli}, {Martin}, {Martin}, {Martynov}, {Mason}, {Massera}, {Masserot},
  {Massinger}, {Masso-Reid}, {Mastrogiovanni}, {Matas}, {Matichard}, {Matone},
  {Mavalvala}, {Mazumder}, {McCann}, {McCarthy}, {McClelland}, {McCormick},
  {McCuller}, {McGuire}, {McIver}, {McManus}, {McRae}, {McWilliams}, {Meacher},
  {Meadors}, {Mehmet}, {Mehta}, {Meidam}, {Melatos}, {Mendell}, {Mercer},
  {Mereni}, {Merilh}, {Merzougui}, {Meshkov}, {Messenger}, {Messick},
  {Metzdorff}, {Meyers}, {Miao}, {Michel}, {Middleton}, {Mikhailov}, {Milano},
  {Miller}, {Miller}, {Millhouse}, {Mills}, {Milovich-Goff}, {Minazzoli},
  {Minenkov}, {Mishkin}, {Mishra}, {Mistry}, {Mitra}, {Mitrofanov},
  {Mitselmakher}, {Mittleman}, {Mo}, {Moffa}, {Mogushi}, {Mohapatra},
  {Montani}, {Moore}, {Moraru}, {Moreno}, {Morisaki}, {Mours}, {Mow-Lowry},
  {Mukherjee}, {Mukherjee}, {Mukherjee}, {Mukund}, {Mullavey}, {Munch},
  {Mu{\~n}iz}, {Muratore}, {Murray}, {Nagar}, {Nardecchia}, {Naticchioni},
  {Nayak}, {Neilson}, {Nelemans}, {Nelson}, {Nery}, {Neunzert}, {Ng}, {Ng},
  {Nguyen}, {Nichols}, {Nissanke}, {Nocera}, {North}, {Nuttall},
  {Obergaulinger}, {Oberling}, {O'Brien}, {O'Dea}, {Ogin}, {Oh}, {Oh}, {Ohme},
  {Ohta}, {Okada}, {Oliver}, {Oppermann}, {Oram}, {O'Reilly}, {Ormiston},
  {Ortega}, {O'Shaughnessy}, {Ossokine}, {Ottaway}, {Overmier}, {Owen}, {Pace},
  {Pagano}, {Page}, {Pai}, {Pai}, {Palamos}, {Palashov}, {Palomba},
  {Pal-Singh}, {Pan}, {Pang}, {Pang}, {Pankow}, {Pannarale}, {Pant},
  {Paoletti}, {Paoli}, {Parida}, {Parker}, {Pascucci}, {Pasqualetti},
  {Passaquieti}, {Passuello}, {Patil}, {Patricelli}, {Pearlstone}, {Pedersen},
  {Pedraza}, {Pedurand}, {Pele}, {Penn}, {Perez}, {Perreca}, {Pfeiffer},
  {Phelps}, {Phukon}, {Piccinni}, {Pichot}, {Piergiovanni}, {Pillant},
  {Pinard}, {Pirello}, {Pitkin}, {Poggiani}, {Pong}, {Ponrathnam}, {Popolizio},
  {Porter}, {Powell}, {Prajapati}, {Prasad}, {Prasai}, {Prasanna}, {Pratten},
  {Prestegard}, {Privitera}, {Prodi}, {Prokhorov}, {Puncken}, {Punturo},
  {Puppo}, {P{\"u}rrer}, {Qi}, {Quetschke}, {Quinonez}, {Quintero},
  {Quitzow-James}, {Raab}, {Radkins}, {Radulescu}, {Raffai}, {Raja}, {Rajan},
  {Rajbhandari}, {Rakhmanov}, {Ramirez}, {Ramos-Buades}, {Rana}, {Rao},
  {Rapagnani}, {Raymond}, {Razzano}, {Read}, {Regimbau}, {Rei}, {Reid},
  {Reitze}, {Ren}, {Ricci}, {Richardson}, {Richardson}, {Ricker}, {Riles},
  {Rizzo}, {Robertson}, {Robie}, {Robinet}, {Rocchi}, {Rolland}, {Rollins},
  {Roma}, {Romanelli}, {Romano}, {Romel}, {Romie}, {Rose}, {Rosi{\'n}ska},
  {Rosofsky}, {Ross}, {Rowan}, {R{\"u}diger}, {Ruggi}, {Rutins}, {Ryan},
  {Sachdev}, {Sadecki}, {Sakellariadou}, {Salconi}, {Saleem}, {Samajdar},
  {Sammut}, {Sanchez}, {Sanchez}, {Sanchis-Gual}, {Sandberg}, {Sanders},
  {Santiago}, {Sarin}, {Sassolas}, {Sathyaprakash}, {Saulson}, {Sauter},
  {Savage}, {Schale}, {Scheel}, {Scheuer}, {Schmidt}, {Schnabel}, {Schofield},
  {Sch{\"o}nbeck}, {Schreiber}, {Schulte}, {Schutz}, {Schwalbe}, {Scott},
  {Scott}, {Seidel}, {Sellers}, {Sengupta}, {Sennett}, {Sentenac}, {Sequino},
  {Sergeev}, {Setyawati}, {Shaddock}, {Shaffer}, {Shahriar}, {Shaner}, {Shao},
  {Sharma}, {Shawhan}, {Shen}, {Shink}, {Shoemaker}, {Shoemaker},
  {ShyamSundar}, {Siellez}, {Sieniawska}, {Sigg}, {Silva}, {Singer}, {Singh},
  {Singhal}, {Sintes}, {Sitmukhambetov}, {Skliris}, {Slagmolen},
  {Slaven-Blair}, {Smith}, {Smith}, {Somala}, {Son}, {Sorazu}, {Sorrentino},
  {Souradeep}, {Sowell}, {Spencer}, {Spera}, {Srivastava}, {Srivastava},
  {Staats}, {Stachie}, {Standke}, {Steer}, {Steinke}, {Steinlechner},
  {Steinlechner}, {Steinmeyer}, {Stevenson}, {Stocks}, {Stone}, {Stops},
  {Strain}, {Stratta}, {Strigin}, {Strunk}, {Sturani}, {Stuver}, {Sudhir},
  {Summerscales}, {Sun}, {Sunil}, {Suresh}, {Sutton}, {Swinkels},
  {Szczepa{\'n}czyk}, {Tacca}, {Tait}, {Talbot}, {Talukder}, {Tanner},
  {T{\'a}pai}, {Taracchini}, {Tasson}, {Taylor}, {Thies}, {Thomas}, {Thomas},
  {Thondapu}, {Thorne}, {Thrane}, {Tiwari}, {Tiwari}, {Tiwari}, {Toland},
  {Tonelli}, {Tornasi}, {Torres-Forn{\'e}}, {Torrie}, {T{\"o}yr{\"a}},
  {Travasso}, {Traylor}, {Tringali}, {Trovato}, {Trozzo}, {Trudeau}, {Tsang},
  {Tse}, {Tso}, {Tsukada}, {Tsuna}, {Tuyenbayev}, {Ueno}, {Ugolini},
  {Unnikrishnan}, {Urban}, {Usman}, {Vahlbruch}, {Vajente}, {Valdes}, {van
  Bakel}, {van Beuzekom}, {van den Brand}, {Van Den Broeck}, {Vander-Hyde},
  {van der Schaaf}, {van Heijningen}, {van Veggel}, {Vardaro}, {Varma}, {Vass},
  {Vas{\'u}th}, {Vecchio}, {Vedovato}, {Veitch}, {Veitch}, {Venkateswara},
  {Venugopalan}, {Verkindt}, {Vetrano}, {Vicer{\'e}}, {Viets}, {Vine}, {Vinet},
  {Vitale}, {Vo}, {Vocca}, {Vorvick}, {Vyatchanin}, {Wade}, {Wade}, {Wade},
  {Walet}, {Walker}, {Wallace}, {Walsh}, {Wang}, {Wang}, {Wang}, {Wang},
  {Wang}, {Ward}, {Warden}, {Warner}, {Was}, {Watchi}, {Weaver}, {Wei},
  {Weinert}, {Weinstein}, {Weiss}, {Wellmann}, {Wen}, {Wessel}, {We{\ss}els},
  {Westhouse}, {Wette}, {Whelan}, {Whiting}, {Whittle}, {Wilken}, {Williams},
  {Williamson}, {Willis}, {Willke}, {Wimmer}, {Winkler}, {Wipf}, {Wittel},
  {Woan}, {Woehler}, {Wofford}, {Worden}, {Wright}, {Wu}, {Wysocki}, {Xiao},
  {Yamamoto}, {Yancey}, {Yang}, {Yap}, {Yazback}, {Yeeles}, {Yu}, {Yu}, {Yuen},
  {Yvert}, {Zadro{\.z}ny}, {Zanolin}, {Zelenova}, {Zendri}, {Zevin}, {Zhang},
  {Zhang}, {Zhang}, {Zhao}, {Zhou}, {Zhou}, {Zhu}, {Zimmerman}, {Zlochower},
  {Zucker}, {Zweizig}, {LIGO Scientific Collaboration}, \& {Virgo
  Collaboration}}]{2019ApJ...882L..24A}
{Abbott}, B.~P., {Abbott}, R., {Abbott}, T.~D., {et~al.} 2019, \apjl, 882, L24

\bibitem[{{Abbott} {et~al.}(2023){Abbott}, {Abbott}, {Acernese}, {Ackley},
  {Adams}, {Adhikari}, {Adhikari}, {Adya}, {Affeldt}, {Agarwal}, {Agathos},
  {Agatsuma}, {Aggarwal}, {Aguiar}, {Aiello}, {Ain}, {Ajith}, {Akutsu}, {de
  Alarc{\'o}n}, {Akcay}, {Albanesi}, {Allocca}, {Altin}, {Amato}, {Anand},
  {Anand}, {Ananyeva}, {Anderson}, {Anderson}, {Ando}, {Andrade}, {Andres},
  {Andri{\'c}}, {Angelova}, {Ansoldi}, {Antelis}, {Antier}, {Antonini},
  {Appert}, {Arai}, {Arai}, {Arai}, {Araki}, {Araya}, {Araya}, {Areeda},
  {Ar{\`e}ne}, {Aritomi}, {Arnaud}, {Arogeti}, {Aronson}, {Arun}, {Asada},
  {Asali}, {Ashton}, {Aso}, {Assiduo}, {Aston}, {Astone}, {Aubin}, {Austin},
  {Babak}, {Badaracco}, {Bader}, {Badger}, {Bae}, {Bae}, {Baer}, {Bagnasco},
  {Bai}, {Baiotti}, {Baird}, {Bajpai}, {Ball}, {Ballardin}, {Ballmer},
  {Balsamo}, {Baltus}, {Banagiri}, {Bankar}, {Barayoga}, {Barbieri}, {Barish},
  {Barker}, {Barneo}, {Barone}, {Barr}, {Barsotti}, {Barsuglia}, {Barta},
  {Bartlett}, {Barton}, {Bartos}, {Bassiri}, {Basti}, {Bawaj}, {Bayley},
  {Baylor}, {Bazzan}, {B{\'e}csy}, {Bedakihale}, {Bejger}, {Belahcene},
  {Benedetto}, {Beniwal}, {Bennett}, {Bentley}, {Benyaala}, {Bergamin},
  {Berger}, {Bernuzzi}, {Berry}, {Bersanetti}, {Bertolini}, {Betzwieser},
  {Beveridge}, {Bhandare}, {Bhardwaj}, {Bhattacharjee}, {Bhaumik}, {Bilenko},
  {Billingsley}, {Bini}, {Birney}, {Birnholtz}, {Biscans}, {Bischi},
  {Biscoveanu}, {Bisht}, {Biswas}, {Bitossi}, {Bizouard}, {Blackburn}, {Blair},
  {Blair}, {Blair}, {Bobba}, {Bode}, {Boer}, {Bogaert}, {Boldrini}, {Bonavena},
  {Bondu}, {Bonilla}, {Bonnand}, {Booker}, {Boom}, {Bork}, {Boschi}, {Bose},
  {Bose}, {Bossilkov}, {Boudart}, {Bouffanais}, {Bozzi}, {Bradaschia}, {Brady},
  {Bramley}, {Branch}, {Branchesi}, {Brandt}, {Brau}, {Breschi}, {Briant},
  {Briggs}, {Brillet}, {Brinkmann}, {Brockill}, {Brooks}, {Brooks}, {Brown},
  {Brunett}, {Bruno}, {Bruntz}, {Bryant}, {Bulik}, {Bulten}, {Buonanno},
  {Buscicchio}, {Buskulic}, {Buy}, {Byer}, {Cadonati}, {Cagnoli}, {Cahillane},
  {Bustillo}, {Callaghan}, {Callister}, {Calloni}, {Cameron}, {Camp}, {Canepa},
  {Canevarolo}, {Cannavacciuolo}, {Cannon}, {Cao}, {Cao}, {Capocasa}, {Capote},
  {Carapella}, {Carbognani}, {Carlin}, {Carney}, {Carpinelli}, {Carrillo},
  {Carullo}, {Carver}, {Diaz}, {Casentini}, {Castaldi}, {Caudill},
  {Cavagli{\`a}}, {Cavalier}, {Cavalieri}, {Ceasar}, {Cella},
  {Cerd{\'a}-Dur{\'a}n}, {Cesarini}, {Chaibi}, {Chakravarti}, {Subrahmanya},
  {Champion}, {Chan}, {Chan}, {Chan}, {Chan}, {Chan}, {Chandra}, {Chanial},
  {Chao}, {Chapman-Bird}, {Charlton}, {Chase}, {Chassande-Mottin},
  {Chatterjee}, {Chatterjee}, {Chatterjee}, {Chaturvedi}, {Chaty},
  {Chatziioannou}, {Chen}, {Chen}, {Chen}, {Chen}, {Chen}, {Chen}, {Chen},
  {Chen}, {Cheng}, {Cheong}, {Cheung}, {Chia}, {Chiadini}, {Chiang},
  {Chiarini}, {Chierici}, {Chincarini}, {Chiofalo}, {Chiummo}, {Cho}, {Cho},
  {Choudhary}, {Choudhary}, {Christensen}, {Chu}, {Chu}, {Chu}, {Chua},
  {Chung}, {Ciani}, {Ciecielag}, {Cie{\'s}lar}, {Cifaldi}, {Ciobanu}, {Ciolfi},
  {Cipriano}, {Cirone}, {Clara}, {Clark}, {Clark}, {Clarke}, {Clearwater},
  {Clesse}, {Cleva}, {Coccia}, {Codazzo}, {Cohadon}, {Cohen}, {Cohen},
  {Colleoni}, {Collette}, {Colombo}, {Colpi}, {Compton}, {Constancio}, {Conti},
  {Cooper}, {Corban}, {Corbitt}, {Cordero-Carri{\'o}n}, {Corezzi}, {Corley},
  {Cornish}, {Corre}, {Corsi}, {Cortese}, {Costa}, {Cotesta}, {Coughlin},
  {Coulon}, {Countryman}, {Cousins}, {Couvares}, {Coward}, {Cowart}, {Coyne},
  {Coyne}, {Creighton}, {Creighton}, {Criswell}, {Croquette}, {Crowder},
  {Cudell}, {Cullen}, {Cumming}, {Cummings}, {Cunningham}, {Cuoco},
  {Cury{\l}o}, {Dabadie}, {Canton}, {Dall'Osso}, {D{\'a}lya}, {Dana},
  {Daneshgaranbajastani}, {D'Angelo}, {Danila}, {Danilishin}, {D'Antonio},
  {Danzmann}, {Darsow-Fromm}, {Dasgupta}, {Datrier}, {Datta}, {Dattilo},
  {Dave}, {Davier}, {Davies}, {Davis}, {Davis}, {Daw}, {Dean}, {Debra},
  {Deenadayalan}, {Degallaix}, {de Laurentis}, {Del{\'e}glise}, {Del Favero},
  {de Lillo}, {de Lillo}, {Del Pozzo}, {Demarchi}, {de Matteis}, {D'Emilio},
  {Demos}, {Dent}, {Depasse}, {de Pietri}, {De Rosa}, {de Rossi}, {Desalvo},
  {de Simone}, {Dhurandhar}, {D{\'\i}az}, {Diaz-Ortiz}, {Didio}, {Dietrich},
  {di Fiore}, {di Fronzo}, {di Giorgio}, {di Giovanni}, {di Giovanni}, {di
  Girolamo}, {di Lieto}, {Ding}, {di Pace}, {di Palma}, {di Renzo},
  {Divakarla}, {Dmitriev}, {Doctor}, {D'Onofrio}, {Donovan}, {Dooley},
  {Doravari}, {Dorrington}, {Drago}, {Driggers}, {Drori}, {Ducoin}, {Dupej},
  {Durante}, {D'Urso}, {Duverne}, {Dwyer}, {Eassa}, {Easter}, {Ebersold},
  {Eckhardt}, {Eddolls}, {Edelman}, {Edo}, {Edy}, {Effler}, {Eguchi},
  {Eichholz}, {Eikenberry}, {Eisenmann}, {Eisenstein}, {Ejlli}, {Engelby},
  {Enomoto}, {Errico}, {Essick}, {Estell{\'e}s}, {Estevez}, {Etienne}, {Etzel},
  {Evans}, {Evans}, {Ewing}, {Fafone}, {Fair}, {Fairhurst}, {Farah}, {Farinon},
  {Farr}, {Farr}, {Farrow}, {Fauchon-Jones}, {Favaro}, {Favata}, {Fays},
  {Fazio}, {Feicht}, {Fejer}, {Fenyvesi}, {Ferguson}, {Fernandez-Galiana},
  {Ferrante}, {Ferreira}, {Fidecaro}, {Figura}, {Fiori}, {Fishbach}, {Fisher},
  {Fittipaldi}, {Fiumara}, {Flaminio}, {Floden}, {Fong}, {Font}, {Fornal},
  {Forsyth}, {Franke}, {Frasca}, {Frasconi}, {Frederick}, {Freed}, {Frei},
  {Freise}, {Frey}, {Fritschel}, {Frolov}, {Fronz{\'e}}, {Fujii}, {Fujikawa},
  {Fukunaga}, {Fukushima}, {Fulda}, {Fyffe}, {Gabbard}, {Gadre}, {Gair},
  {Gais}, {Galaudage}, {Gamba}, {Ganapathy}, {Ganguly}, {Gao}, {Gaonkar},
  {Garaventa}, {Garc{\'\i}a}, {Garc{\'\i}a-N{\'u}{\~n}ez},
  {Garc{\'\i}a-Quir{\'o}s}, {Garufi}, {Gateley}, {Gaudio}, {Gayathri}, {Ge},
  {Gemme}, {Gennai}, {George}, {George}, {Gerberding}, {Gergely}, {Gewecke},
  {Ghonge}, {Ghosh}, {Ghosh}, {Ghosh}, {Ghosh}, {Giacomazzo}, {Giacoppo},
  {Giaime}, {Giardina}, {Gibson}, {Gier}, {Giesler}, {Giri}, {Gissi},
  {Glanzer}, {Gleckl}, {Godwin}, {Golomb}, {Goetz}, {Goetz}, {Gohlke},
  {Goncharov}, {Gonz{\'a}lez}, {Gopakumar}, {Gosselin}, {Gouaty}, {Gould},
  {Grace}, {Grado}, {Granata}, {Granata}, {Grant}, {Gras}, {Grassia}, {Gray},
  {Gray}, {Greco}, {Green}, {Green}, {Gretarsson}, {Gretarsson}, {Griffith},
  {Griffiths}, {Griggs}, {Grignani}, {Grimaldi}, {Grimm}, {Grote}, {Grunewald},
  {Gruning}, {Guerra}, {Guidi}, {Guimaraes}, {Guix{\'e}}, {Gulati}, {Guo},
  {Guo}, {Gupta}, {Gupta}, {Gupta}, {Gustafson}, {Gustafson}, {Guzman}, {Ha},
  {Haegel}, {Hagiwara}, {Haino}, {Halim}, {Hall}, {Hamilton}, {Hammond}, {Han},
  {Haney}, {Hanks}, {Hanna}, {Hannam}, {Hannuksela}, {Hansen}, {Hansen},
  {Hanson}, {Harder}, {Hardwick}, {Haris}, {Harms}, {Harry}, {Harry},
  {Hartwig}, {Hasegawa}, {Haskell}, {Hasskew}, {Haster}, {Hattori}, {Haughian},
  {Hayakawa}, {Hayama}, {Hayes}, {Healy}, {Heidmann}, {Heidt}, {Heintze},
  {Heinze}, {Heinzel}, {Heitmann}, {Hellman}, {Hello}, {Helmling-Cornell},
  {Hemming}, {Hendry}, {Heng}, {Hennes}, {Hennig}, {Hennig}, {Hernandez},
  {Vivanco}, {Heurs}, {Hild}, {Hill}, {Himemoto}, {Hines}, {Hiranuma},
  {Hirata}, {Hirose}, {Hochheim}, {Hofman}, {Hohmann}, {Holcomb}, {Holland},
  {Hollows}, {Holmes}, {Holt}, {Holz}, {Hong}, {Hopkins}, {Hough}, {Hourihane},
  {Howell}, {Hoy}, {Hoyland}, {Hreibi}, {Hsieh}, {Hsu}, {Huang}, {Huang},
  {Huang}, {Huang}, {Huang}, {Huang}, {H{\"u}bner}, {Huddart}, {Hughey}, {Hui},
  {Hui}, {Husa}, {Huttner}, {Huxford}, {Huynh-Dinh}, {Ide}, {Idzkowski},
  {Iess}, {Ikenoue}, {Imam}, {Inayoshi}, {Ingram}, {Inoue}, {Ioka}, {Isi},
  {Isleif}, {Ito}, {Itoh}, {Iyer}, {Izumi}, {Jaberianhamedan}, {Jacqmin},
  {Jadhav}, {Jadhav}, {James}, {Jan}, {Jani}, {Janquart}, {Janssens},
  {Janthalur}, {Jaranowski}, {Jariwala}, {Jaume}, {Jenkins}, {Jenner}, {Jeon},
  {Jeunon}, {Jia}, {Jin}, {Johns}, {Jones}, {Jones}, {Jones}, {Jones}, {Jones},
  {Jonker}, {Ju}, {Jung}, {Jung}, {Junker}, {Juste}, {Kaihotsu}, {Kajita},
  {Kakizaki}, {Kalaghatgi}, {Kalogera}, {Kamai}, {Kamiizumi}, {Kanda},
  {Kandhasamy}, {Kang}, {Kanner}, {Kao}, {Kapadia}, {Kapasi}, {Karat},
  {Karathanasis}, {Karki}, {Kashyap}, {Kasprzack}, {Kastaun}, {Katsanevas},
  {Katsavounidis}, {Katzman}, {Kaur}, {Kawabe}, {Kawaguchi}, {Kawai},
  {Kawasaki}, {K{\'e}f{\'e}lian}, {Keitel}, {Key}, {Khadka}, {Khalili}, {Khan},
  {Khazanov}, {Khetan}, {Khursheed}, {Kijbunchoo}, {Kim}, {Kim}, {Kim}, {Kim},
  {Kim}, {Kim}, {Kimball}, {Kimura}, {Kinley-Hanlon}, {Kirchhoff}, {Kissel},
  {Kita}, {Kitazawa}, {Kleybolte}, {Klimenko}, {Knee}, {Knowles}, {Knyazev},
  {Koch}, {Koekoek}, {Kojima}, {Kokeyama}, {Koley}, {Kolitsidou}, {Kolstein},
  {Komori}, {Kondrashov}, {Kong}, {Kontos}, {Koper}, {Korobko}, {Kotake},
  {Kovalam}, {Kozak}, {Kozakai}, {Kozu}, {Kringel}, {Krishnendu}, {Kr{\'o}lak},
  {Kuehn}, {Kuei}, {Kuijer}, {Kulkarni}, {Kumar}, {Kumar}, {Kumar}, {Kumar},
  {Kume}, {Kuns}, {Kuo}, {Kuo}, {Kuromiya}, {Kuroyanagi}, {Kusayanagi},
  {Kuwahara}, {Kwak}, {Lagabbe}, {Laghi}, {Lalande}, {Lam}, {Lamberts},
  {Landry}, {Landry}, {Lane}, {Lang}, {Lange}, {Lantz}, {La Rosa},
  {Lartaux-Vollard}, {Lasky}, {Laxen}, {Lazzarini}, {Lazzaro}, {Leaci},
  {Leavey}, {Lecoeuche}, {Lee}, {Lee}, {Lee}, {Lee}, {Lee}, {Lee}, {Lehmann},
  {Lema{\^\i}tre}, {Leonardi}, {Leroy}, {Letendre}, {Levesque}, {Levin},
  {Leviton}, {Leyde}, {Li}, {Li}, {Li}, {Li}, {Li}, {Li}, {Lin}, {Lin}, {Lin},
  {Lin}, {Lin}, {Linde}, {Linker}, {Linley}, {Littenberg}, {Liu}, {Liu}, {Liu},
  {Liu}, {Llamas}, {Llorens-Monteagudo}, {Lo}, {Lockwood}, {Loh}, {London},
  {Longo}, {Lopez}, {Portilla}, {Lorenzini}, {Loriette}, {Lormand}, {Losurdo},
  {Lott}, {Lough}, {Lousto}, {Lovelace}, {Lucaccioni}, {L{\"u}ck}, {Lumaca},
  {Lundgren}, {Luo}, {Lynam}, {Macas}, {Macinnis}, {MacLeod}, {MacMillan},
  {Macquet}, {Hernandez}, {Magazz{\`u}}, {Magee}, {Maggiore}, {Magnozzi},
  {Mahesh}, {Majorana}, {Makarem}, {Maksimovic}, {Maliakal}, {Malik}, {Man},
  {Mandic}, {Mangano}, {Mango}, {Mansell}, {Manske}, {Mantovani}, {Mapelli},
  {Marchesoni}, {Marchio}, {Marion}, {Mark}, {M{\'a}rka}, {M{\'a}rka},
  {Markakis}, {Markosyan}, {Markowitz}, {Maros}, {Marquina}, {Marsat},
  {Martelli}, {Martin}, {Martin}, {Martinez}, {Martinez}, {Martinez},
  {Martinovic}, {Martynov}, {Marx}, {Masalehdan}, {Mason}, {Massera},
  {Masserot}, {Massinger}, {Masso-Reid}, {Mastrogiovanni}, {Matas},
  {Mateu-Lucena}, {Matichard}, {Matiushechkina}, {Mavalvala}, {McCann},
  {McCarthy}, {McClelland}, {McClincy}, {McCormick}, {McCuller}, {McGhee},
  {McGuire}, {McIsaac}, {McIver}, {McRae}, {McWilliams}, {Meacher}, {Mehmet},
  {Mehta}, {Meijer}, {Melatos}, {Melchor}, {Mendell}, {Menendez-Vazquez},
  {Menoni}, {Mercer}, {Mereni}, {Merfeld}, {Merilh}, {Merritt}, {Merzougui},
  {Meshkov}, {Messenger}, {Messick}, {Meyers}, {Meylahn}, {Mhaske}, {Miani},
  {Miao}, {Michaloliakos}, {Michel}, {Michimura}, {Middleton}, {Milano},
  {Miller}, {Miller}, {Miller}, {Miller}, {Millhouse}, {Mills}, {Milotti},
  {Minazzoli}, {Minenkov}, {Mio}, {Mir}, {Miravet-Ten{\'e}s}, {Mishra},
  {Mishra}, {Mistry}, {Mitra}, {Mitrofanov}, {Mitselmakher}, {Mittleman},
  {Miyakawa}, {Miyamoto}, {Miyazaki}, {Miyo}, {Miyoki}, {Mo}, {Modafferi},
  {Moguel}, {Mogushi}, {Mohapatra}, {Mohite}, {Molina}, {Molina-Ruiz},
  {Mondin}, {Montani}, {Moore}, {Moraru}, {Morawski}, {More}, {Moreno},
  {Moreno}, {Mori}, {Morisaki}, {Moriwaki}, {Morr{\'a}s}, {Mours}, {Mow-Lowry},
  {Mozzon}, {Muciaccia}, {Mukherjee}, {Mukherjee}, {Mukherjee}, {Mukherjee},
  {Mukherjee}, {Mukund}, {Mullavey}, {Munch}, {Mu{\~n}iz}, {Murray},
  {Musenich}, {Muusse}, {Nadji}, {Nagano}, {Nagano}, {Nagar}, {Nakamura},
  {Nakano}, {Nakano}, {Nakashima}, {Nakayama}, {Napolano}, {Nardecchia},
  {Narikawa}, {Naticchioni}, {Nayak}, {Nayak}, {Negishi}, {Neil}, {Neilson},
  {Nelemans}, {Nelson}, {Nery}, {Neubauer}, {Neunzert}, {Ng}, {Ng}, {Nguyen},
  {Nguyen}, {Nguyen}, {Quynh}, {Ni}, {Nichols}, {Nishizawa}, {Nissanke},
  {Nitoglia}, {Nocera}, {Norman}, {North}, {Nozaki}, {Siles}, {Nuttall},
  {Oberling}, {O'Brien}, {Obuchi}, {O'Dell}, {Oelker}, {Ogaki}, {Oganesyan},
  {Oh}, {Oh}, {Oh}, {Ohashi}, {Ohishi}, {Ohkawa}, {Ohme}, {Ohta}, {Okada},
  {Okutani}, {Okutomi}, {Olivetto}, {Oohara}, {Ooi}, {Oram}, {O'Reilly},
  {Ormiston}, {Ormsby}, {Ortega}, {O'Shaughnessy}, {O'Shea}, {Oshino},
  {Ossokine}, {Osthelder}, {Otabe}, {Ottaway}, {Overmier}, {Pace}, {Pagano},
  {Page}, {Pagliaroli}, {Pai}, {Pai}, {Palamos}, {Palashov}, {Palomba}, {Pan},
  {Pan}, {Panda}, {Pang}, {Pang}, {Pankow}, {Pannarale}, {Pant}, {Panther},
  {Paoletti}, {Paoli}, {Paolone}, {Parisi}, {Park}, {Park}, {Parker},
  {Pascucci}, {Pasqualetti}, {Passaquieti}, {Passuello}, {Patel}, {Pathak},
  {Patricelli}, {Patron}, {Paul}, {Payne}, {Pedraza}, {Pegoraro}, {Pele},
  {Arellano}, {Penn}, {Perego}, {Pereira}, {Pereira}, {Perez}, {P{\'e}rigois},
  {Perkins}, {Perreca}, {Perri{\`e}s}, {Petermann}, {Petterson}, {Pfeiffer},
  {Pham}, {Phukon}, {Piccinni}, {Pichot}, {Piendibene}, {Piergiovanni},
  {Pierini}, {Pierro}, {Pillant}, {Pillas}, {Pilo}, {Pinard}, {Pinto}, {Pinto},
  {Piotrzkowski}, {Piotrzkowski}, {Pirello}, {Pitkin}, {Placidi}, {Planas},
  {Plastino}, {Pluchar}, {Poggiani}, {Polini}, {Pong}, {Ponrathnam},
  {Popolizio}, {Porter}, {Poulton}, {Powell}, {Pracchia}, {Pradier},
  {Prajapati}, {Prasai}, {Prasanna}, {Pratten}, {Principe}, {Prodi},
  {Prokhorov}, {Prosposito}, {Prudenzi}, {Puecher}, {Punturo}, {Puosi},
  {Puppo}, {P{\"u}rrer}, {Qi}, {Quetschke}, {Quitzow-James}, {Raab},
  {Raaijmakers}, {Radkins}, {Radulesco}, {Raffai}, {Rail}, {Raja}, {Rajan},
  {Ramirez}, {Ramirez}, {Ramos-Buades}, {Rana}, {Rapagnani}, {Rapol}, {Ray},
  {Raymond}, {Raza}, {Razzano}, {Read}, {Rees}, {Regimbau}, {Rei}, {Reid},
  {Reid}, {Reitze}, {Relton}, {Renzini}, {Rettegno}, {Reza}, {Rezac}, {Ricci},
  {Richards}, {Richardson}, {Richardson}, {Riemenschneider}, {Riles},
  {Rinaldi}, {Rink}, {Rizzo}, {Robertson}, {Robie}, {Robinet}, {Rocchi},
  {Rodriguez}, {Rolland}, {Rollins}, {Romanelli}, {Romano}, {Romel},
  {Romero-Rodr{\'\i}guez}, {Romero-Shaw}, {Romie}, {Ronchini}, {Rosa}, {Rose},
  {Rosi{\'n}ska}, {Ross}, {Rowan}, {Rowlinson}, {Roy}, {Roy}, {Roy}, {Rozza},
  {Ruggi}, {Ryan}, {Sachdev}, {Sadecki}, {Sadiq}, {Sago}, {Saito}, {Saito},
  {Sakai}, {Sakai}, {Sakellariadou}, {Sakuno}, {Salafia}, {Salconi}, {Saleem},
  {Salemi}, {Samajdar}, {Sanchez}, {Sanchez}, {Sanchez}, {Sanchis-Gual},
  {Sanders}, {Sanuy}, {Saravanan}, {Sarin}, {Sassolas}, {Satari},
  {Sathyaprakash}, {Sato}, {Sato}, {Sauter}, {Savage}, {Sawada}, {Sawant},
  {Sawant}, {Sayah}, {Schaetzl}, {Scheel}, {Scheuer}, {Schiworski}, {Schmidt},
  {Schmidt}, {Schnabel}, {Schneewind}, {Schofield}, {Sch{\"o}nbeck}, {Schulte},
  {Schutz}, {Schwartz}, {Scott}, {Scott}, {Seglar-Arroyo}, {Sekiguchi},
  {Sekiguchi}, {Sellers}, {Sengupta}, {Sentenac}, {Seo}, {Sequino}, {Sergeev},
  {Setyawati}, {Shaffer}, {Shahriar}, {Shams}, {Shao}, {Sharma}, {Sharma},
  {Shawhan}, {Shcheblanov}, {Shibagaki}, {Shikauchi}, {Shimizu}, {Shimoda},
  {Shimode}, {Shinkai}, {Shishido}, {Shoda}, {Shoemaker}, {Shoemaker},
  {Shyamsundar}, {Sieniawska}, {Sigg}, {Singer}, {Singh}, {Singh}, {Singha},
  {Sintes}, {Sipala}, {Skliris}, {Slagmolen}, {Slaven-Blair}, {Smetana},
  {Smith}, {Smith}, {Soldateschi}, {Somala}, {Somiya}, {Son}, {Soni}, {Soni},
  {Sordini}, {Sorrentino}, {Sorrentino}, {Sotani}, {Soulard}, {Souradeep},
  {Sowell}, {Spagnuolo}, {Spencer}, {Spera}, {Srinivasan}, {Srivastava},
  {Srivastava}, {Staats}, {Stachie}, {Steer}, {Steinhoff}, {Steinlechner},
  {Steinlechner}, {Stevenson}, {Stops}, {Stover}, {Strain}, {Strang},
  {Stratta}, {Strunk}, {Sturani}, {Stuver}, {Sudhagar}, {Sudhir}, {Sugimoto},
  {Suh}, {Sullivan}, {Summerscales}, {Sun}, {Sun}, {Sunil}, {Sur}, {Suresh},
  {Sutton}, {Suzuki}, {Suzuki}, {Swinkels}, {Szczepa{\'n}czyk}, {Szewczyk},
  {Tacca}, {Tagoshi}, {Tait}, {Takahashi}, {Takahashi}, {Takamori}, {Takano},
  {Takeda}, {Takeda}, {Talbot}, {Talbot}, {Tanaka}, {Tanaka}, {Tanaka},
  {Tanaka}, {Tanaka}, {Tanasijczuk}, {Tanioka}, {Tanner}, {Tao}, {Tao},
  {Mart{\'\i}n}, {Taranto}, {Tasson}, {Telada}, {Tenorio}, {Terhune},
  {Terkowski}, {Thirugnanasambandam}, {Thomas}, {Thomas}, {Thomas}, {Thompson},
  {Thondapu}, {Thorne}, {Thrane}, {Tiwari}, {Tiwari}, {Tiwari}, {Toivonen},
  {Toland}, {Tolley}, {Tomaru}, {Tomigami}, {Tomura}, {Tonelli},
  {Torres-Forn{\'e}}, {Torrie}, {E Melo}, {T{\"o}yr{\"a}}, {Trapananti},
  {Travasso}, {Traylor}, {Trevor}, {Tringali}, {Tripathee}, {Troiano},
  {Trovato}, {Trozzo}, {Trudeau}, {Tsai}, {Tsai}, {Tsang}, {Tsang}, {Tsao},
  {Tse}, {Tso}, {Tsubono}, {Tsuchida}, {Tsukada}, {Tsuna}, {Tsutsui},
  {Tsuzuki}, {Turbang}, {Turconi}, {Tuyenbayev}, {Ubhi}, {Uchikata},
  {Uchiyama}, {Udall}, {Ueda}, {Uehara}, {Ueno}, {Ueshima}, {Unnikrishnan},
  {Uraguchi}, {Urban}, {Ushiba}, {Utina}, {Vahlbruch}, {Vajente}, {Vajpeyi},
  {Valdes}, {Valentini}, {Valsan}, {van Bakel}, {van Beuzekom}, {van den
  Brand}, {van den Broeck}, {Vander-Hyde}, {van der Schaaf}, {van Heijningen},
  {Vanosky}, {van Putten}, {van Remortel}, {Vardaro}, {Vargas}, {Varma},
  {Vas{\'u}th}, {Vecchio}, {Vedovato}, {Veitch}, {Veitch}, {Venneberg},
  {Venugopalan}, {Verkindt}, {Verma}, {Verma}, {Veske}, {Vetrano},
  {Vicer{\'e}}, {Vidyant}, {Viets}, {Vijaykumar}, {Villa-Ortega}, {Vinet},
  {Virtuoso}, {Vitale}, {Vo}, {Vocca}, {von Reis}, {von Wrangel}, {Vorvick},
  {Vyatchanin}, {Wade}, {Wade}, {Wagner}, {Walet}, {Walker}, {Wallace},
  {Wallace}, {Walsh}, {Wang}, {Wang}, {Wang}, {Ward}, {Warner}, {Was},
  {Washimi}, {Washington}, {Watchi}, {Weaver}, {Webster}, {Weinert},
  {Weinstein}, {Weiss}, {Weller}, {Wellmann}, {Wen}, {We{\ss}els}, {Wette},
  {Whelan}, {White}, {Whiting}, {Whittle}, {Wilken}, {Williams}, {Williams},
  {Williamson}, {Willis}, {Willke}, {Wilson}, {Winkler}, {Wipf}, {Wlodarczyk},
  {Woan}, {Woehler}, {Wofford}, {Wong}, {Wu}, {Wu}, {Wu}, {Wu}, {Wysocki},
  {Xiao}, {Xu}, {Yamada}, {Yamamoto}, {Yamamoto}, {Yamamoto}, {Yamamoto},
  {Yamashita}, {Yamazaki}, {Yang}, {Yang}, {Yang}, {Yang}, {Yang}, {Yap},
  {Yeeles}, {Yelikar}, {Ying}, {Yokogawa}, {Yokoyama}, {Yokozawa}, {Yoo},
  {Yoshioka}, {Yu}, {Yu}, {Yuzurihara}, {Zadro{\.z}ny}, {Zanolin}, {Zeidler},
  {Zelenova}, {Zendri}, {Zevin}, {Zhan}, {Zhang}, {Zhang}, {Zhang}, {Zhang},
  {Zhang}, {Zhao}, {Zhao}, {Zhao}, {Zhao}, {Zheng}, {Zhou}, {Zhou}, {Zhu},
  {Zhu}, {Zimmerman}, {Zlochower}, {Zucker}, {Zweizig}, {LIGO Scientific
  Collaboration}, {VIRGO Collaboration}, \& {KAGRA
  Collaboration}}]{2023PhRvX..13a1048A}
{Abbott}, R., {Abbott}, T.~D., {Acernese}, F., {et~al.} 2023, Physical Review
  X, 13, 011048

\bibitem[{{Akira Rocha} {et~al.}(2024){Akira Rocha}, {Kalogera}, {Doctor},
  {Andrews}, {Sun}, {Gossage}, {Bavera}, {Fragos}, {Kovlakas}, {Kruckow},
  {Misra}, {Srivastava}, {Xing}, \& {Zapartas}}]{2024arXiv240307172A}
{Akira Rocha}, K., {Kalogera}, V., {Doctor}, Z., {et~al.} 2024, arXiv e-prints,
  arXiv:2403.07172

\bibitem[{{Bardeen}(1970)}]{1970Natur.226...64B}
{Bardeen}, J.~M. 1970, \nat, 226, 64

\bibitem[{{Batta} {et~al.}(2017){Batta}, {Ramirez-Ruiz}, \&
  {Fryer}}]{2017ApJ...846L..15B}
{Batta}, A., {Ramirez-Ruiz}, E., \& {Fryer}, C. 2017, \apjl, 846, L15

\bibitem[{{Bavera} {et~al.}(2020){Bavera}, {Fragos}, {Qin}, {Zapartas},
  {Neijssel}, {Mandel}, {Batta}, {Gaebel}, {Kimball}, \&
  {Stevenson}}]{2020A&A...635A..97B}
{Bavera}, S.~S., {Fragos}, T., {Qin}, Y., {et~al.} 2020, \aap, 635, A97

\bibitem[{{Bavera} {et~al.}(2023){Bavera}, {Fragos}, {Zapartas}, {Andrews},
  {Kalogera}, {Berry}, {Kruckow}, {Dotter}, {Kovlakas}, {Misra}, {Rocha},
  {Srivastava}, {Sun}, \& {Xing}}]{2023NatAs...7.1090B}
{Bavera}, S.~S., {Fragos}, T., {Zapartas}, E., {et~al.} 2023, Nature Astronomy,
  7, 1090

\bibitem[{{Bavera} {et~al.}(2021){Bavera}, {Fragos}, {Zevin}, {Berry},
  {Marchant}, {Andrews}, {Coughlin}, {Dotter}, {Kovlakas}, {Misra},
  {Serra-Perez}, {Qin}, {Rocha}, {Rom{\'a}n-Garza}, {Tran}, \&
  {Zapartas}}]{2021A&A...647A.153B}
{Bavera}, S.~S., {Fragos}, T., {Zevin}, M., {et~al.} 2021, \aap, 647, A153

\bibitem[{{Beck} {et~al.}(2012){Beck}, {Montalban}, {Kallinger}, {De Ridder},
  {Aerts}, {Garc{\'\i}a}, {Hekker}, {Dupret}, {Mosser}, {Eggenberger},
  {Stello}, {Elsworth}, {Frandsen}, {Carrier}, {Hillen}, {Gruberbauer},
  {Christensen-Dalsgaard}, {Miglio}, {Valentini}, {Bedding}, {Kjeldsen},
  {Girouard}, {Hall}, \& {Ibrahim}}]{2012Natur.481...55B}
{Beck}, P.~G., {Montalban}, J., {Kallinger}, T., {et~al.} 2012, \nat, 481, 55

\bibitem[{{Belczynski} {et~al.}(2020){Belczynski}, {Klencki}, {Fields},
  {Olejak}, {Berti}, {Meynet}, {Fryer}, {Holz}, {O'Shaughnessy}, {Brown},
  {Bulik}, {Leung}, {Nomoto}, {Madau}, {Hirschi}, {Kaiser}, {Jones}, {Mondal},
  {Chruslinska}, {Drozda}, {Gerosa}, {Doctor}, {Giersz}, {Ekstrom}, {Georgy},
  {Askar}, {Baibhav}, {Wysocki}, {Natan}, {Farr}, {Wiktorowicz}, {Coleman
  Miller}, {Farr}, \& {Lasota}}]{2020A&A...636A.104B}
{Belczynski}, K., {Klencki}, J., {Fields}, C.~E., {et~al.} 2020, \aap, 636,
  A104

\bibitem[{{Bondi} \& {Hoyle}(1944)}]{1944MNRAS.104..273B}
{Bondi}, H. \& {Hoyle}, F. 1944, \mnras, 104, 273

\bibitem[{{Briel} {et~al.}(2023){Briel}, {Stevance}, \&
  {Eldridge}}]{2023MNRAS.520.5724B}
{Briel}, M.~M., {Stevance}, H.~F., \& {Eldridge}, J.~J. 2023, \mnras, 520, 5724

\bibitem[{{Crowther} {et~al.}(2010){Crowther}, {Barnard}, {Carpano}, {Clark},
  {Dhillon}, \& {Pollock}}]{2010MNRAS.403L..41C}
{Crowther}, P.~A., {Barnard}, R., {Carpano}, S., {et~al.} 2010, \mnras, 403,
  L41

\bibitem[{{Dai} {et~al.}(2018){Dai}, {McKinney}, {Roth}, {Ramirez-Ruiz}, \&
  {Miller}}]{Dai18}
{Dai}, L., {McKinney}, J.~C., {Roth}, N., {Ramirez-Ruiz}, E., \& {Miller},
  M.~C. 2018, \apj, 859, L20

\bibitem[{{de Jager} {et~al.}(1988){de Jager}, {Nieuwenhuijzen}, \& {van der
  Hucht}}]{1988A&AS...72..259D}
{de Jager}, C., {Nieuwenhuijzen}, H., \& {van der Hucht}, K.~A. 1988, \aaps,
  72, 259

\bibitem[{{Deheuvels} {et~al.}(2014){Deheuvels}, {Do{\u{g}}an}, {Goupil},
  {Appourchaux}, {Benomar}, {Bruntt}, {Campante}, {Casagrande}, {Ceillier},
  {Davies}, {De Cat}, {Fu}, {Garc{\'\i}a}, {Lobel}, {Mosser}, {Reese},
  {Regulo}, {Schou}, {Stahn}, {Thygesen}, {Yang}, {Chaplin},
  {Christensen-Dalsgaard}, {Eggenberger}, {Gizon}, {Mathis},
  {Molenda-{\.Z}akowicz}, \& {Pinsonneault}}]{2014A&A...564A..27D}
{Deheuvels}, S., {Do{\u{g}}an}, G., {Goupil}, M.~J., {et~al.} 2014, \aap, 564,
  A27

\bibitem[{{Dorozsmai} \& {Toonen}(2024)}]{2024MNRAS.tmp..142D}
{Dorozsmai}, A. \& {Toonen}, S. 2024, \mnras [\eprint[arXiv]{2207.08837}]

\bibitem[{{Draghis} {et~al.}(2023){Draghis}, {Miller}, {Costantini}, {Gallo},
  {Reynolds}, {Tomsick}, \& {Zoghbi}}]{2023arXiv231116225D}
{Draghis}, P.~A., {Miller}, J.~M., {Costantini}, E., {et~al.} 2023, arXiv
  e-prints, arXiv:2311.16225

\bibitem[{{Eggenberger} {et~al.}(2005){Eggenberger}, {Maeder}, \&
  {Meynet}}]{2005A&A...440L...9E}
{Eggenberger}, P., {Maeder}, A., \& {Meynet}, G. 2005, \aap, 440, L9

\bibitem[{{Eggenberger} {et~al.}(2022){Eggenberger}, {Moyano}, \& {den
  Hartogh}}]{2022A&A...664L..16E}
{Eggenberger}, P., {Moyano}, F.~D., \& {den Hartogh}, J.~W. 2022, \aap, 664,
  L16

\bibitem[{{Falanga} {et~al.}(2021){Falanga}, {Bakala}, {La Placa}, {De Falco},
  {De Rosa}, \& {Stella}}]{2021MNRAS.504.3424F}
{Falanga}, M., {Bakala}, P., {La Placa}, R., {et~al.} 2021, \mnras, 504, 3424

\bibitem[{{Farrell} {et~al.}(2022){Farrell}, {Groh}, {Meynet}, \&
  {Eldridge}}]{2022MNRAS.512.4116F}
{Farrell}, E., {Groh}, J.~H., {Meynet}, G., \& {Eldridge}, J.~J. 2022, \mnras,
  512, 4116

\bibitem[{{Fishbach} \& {Kalogera}(2022)}]{2022ApJ...929L..26F}
{Fishbach}, M. \& {Kalogera}, V. 2022, \apjl, 929, L26

\bibitem[{{Fragos} {et~al.}(2023){Fragos}, {Andrews}, {Bavera}, {Berry},
  {Coughlin}, {Dotter}, {Giri}, {Kalogera}, {Katsaggelos}, {Kovlakas},
  {Lalvani}, {Misra}, {Srivastava}, {Qin}, {Rocha}, {Rom{\'a}n-Garza}, {Serra},
  {Stahle}, {Sun}, {Teng}, {Trajcevski}, {Tran}, {Xing}, {Zapartas}, \&
  {Zevin}}]{2023ApJS..264...45F}
{Fragos}, T., {Andrews}, J.~J., {Bavera}, S.~S., {et~al.} 2023, \apjs, 264, 45

\bibitem[{{Fragos} \& {McClintock}(2015)}]{2015ApJ...800...17F}
{Fragos}, T. \& {McClintock}, J.~E. 2015, \apj, 800, 17

\bibitem[{{Fuller} \& {Ma}(2019)}]{2019ApJ...881L...1F}
{Fuller}, J. \& {Ma}, L. 2019, \apjl, 881, L1

\bibitem[{{Fuller} {et~al.}(2019){Fuller}, {Piro}, \&
  {Jermyn}}]{2019MNRAS.485.3661F}
{Fuller}, J., {Piro}, A.~L., \& {Jermyn}, A.~S. 2019, \mnras, 485, 3661

\bibitem[{{Gallegos-Garcia} {et~al.}(2022){Gallegos-Garcia}, {Fishbach},
  {Kalogera}, {L Berry}, \& {Doctor}}]{2022ApJ...938L..19G}
{Gallegos-Garcia}, M., {Fishbach}, M., {Kalogera}, V., {L Berry}, C.~P., \&
  {Doctor}, Z. 2022, \apjl, 938, L19

\bibitem[{{Gammie} {et~al.}(2004){Gammie}, {Shapiro}, \&
  {McKinney}}]{Gammie.2004}
{Gammie}, C.~F., {Shapiro}, S.~L., \& {McKinney}, J.~C. 2004, \apj, 602, 312

\bibitem[{{Gehan} {et~al.}(2018){Gehan}, {Mosser}, {Michel}, {Samadi}, \&
  {Kallinger}}]{2018A&A...616A..24G}
{Gehan}, C., {Mosser}, B., {Michel}, E., {Samadi}, R., \& {Kallinger}, T. 2018,
  \aap, 616, A24

\bibitem[{{Gomez} \& {Grindlay}(2021)}]{2021ApJ...913...48G}
{Gomez}, S. \& {Grindlay}, J.~E. 2021, \apj, 913, 48

\bibitem[{Hirai \& Mandel(2021)}]{hirai_mandel_2021}
Hirai, R. \& Mandel, I. 2021, Publications of the Astronomical Society of
  Australia, 38, e056

\bibitem[{{Hobbs} {et~al.}(2005){Hobbs}, {Lorimer}, {Lyne}, \&
  {Kramer}}]{2005MNRAS.360..974H}
{Hobbs}, G., {Lorimer}, D.~R., {Lyne}, A.~G., \& {Kramer}, M. 2005, \mnras,
  360, 974

\bibitem[{{Jermyn} {et~al.}(2023){Jermyn}, {Bauer}, {Schwab}, {Farmer}, {Ball},
  {Bellinger}, {Dotter}, {Joyce}, {Marchant}, {Mombarg}, {Wolf}, {Sunny Wong},
  {Cinquegrana}, {Farrell}, {Smolec}, {Thoul}, {Cantiello}, {Herwig}, {Toloza},
  {Bildsten}, {Townsend}, \& {Timmes}}]{2023ApJS..265...15J}
{Jermyn}, A.~S., {Bauer}, E.~B., {Schwab}, J., {et~al.} 2023, \apjs, 265, 15

\bibitem[{{King} {et~al.}(2001){King}, {Davies}, {Ward}, {Fabbiano}, \&
  {Elvis}}]{2001ApJ...552L.109K}
{King}, A.~R., {Davies}, M.~B., {Ward}, M.~J., {Fabbiano}, G., \& {Elvis}, M.
  2001, \apjl, 552, L109

\bibitem[{{Kovlakas} {et~al.}(2022){Kovlakas}, {Fragos}, {Schaerer}, \&
  {Mesinger}}]{2022A&A...665A..28K}
{Kovlakas}, K., {Fragos}, T., {Schaerer}, D., \& {Mesinger}, A. 2022, \aap,
  665, A28

\bibitem[{{Kovlakas} {et~al.}(2020){Kovlakas}, {Zezas}, {Andrews}, {Basu-Zych},
  {Fragos}, {Hornschemeier}, {Lehmer}, \& {Ptak}}]{2020MNRAS.498.4790K}
{Kovlakas}, K., {Zezas}, A., {Andrews}, J.~J., {et~al.} 2020, \mnras, 498, 4790

\bibitem[{{Kroupa}(2001)}]{2001MNRAS.322..231K}
{Kroupa}, P. 2001, \mnras, 322, 231

\bibitem[{{Kwan} {et~al.}(in prep.){Kwan}, {Dai}, {Kan}, {Fragos}, {Middleton},
  \& {Xing}}]{Kwan.prep}
{Kwan}, T.~M., {Dai}, L., {Kan}, Z. C.~K., {et~al.} in prep.

\bibitem[{{Laycock} {et~al.}(2015){Laycock}, {Maccarone}, \&
  {Christodoulou}}]{2015MNRAS.452L..31L}
{Laycock}, S. G.~T., {Maccarone}, T.~J., \& {Christodoulou}, D.~M. 2015,
  \mnras, 452, L31

\bibitem[{{Liotine} {et~al.}(2023){Liotine}, {Zevin}, {Berry}, {Doctor}, \&
  {Kalogera}}]{2023ApJ...946....4L}
{Liotine}, C., {Zevin}, M., {Berry}, C. P.~L., {Doctor}, Z., \& {Kalogera}, V.
  2023, \apj, 946, 4

\bibitem[{{Liu} {et~al.}(2008){Liu}, {McClintock}, {Narayan}, {Davis}, \&
  {Orosz}}]{2008ApJ...679L..37L}
{Liu}, J., {McClintock}, J.~E., {Narayan}, R., {Davis}, S.~W., \& {Orosz},
  J.~A. 2008, \apjl, 679, L37

\bibitem[{{Liu} {et~al.}(2010){Liu}, {McClintock}, {Narayan}, {Davis}, \&
  {Orosz}}]{2010ApJ...719L.109L}
{Liu}, J., {McClintock}, J.~E., {Narayan}, R., {Davis}, S.~W., \& {Orosz},
  J.~A. 2010, \apjl, 719, L109

\bibitem[{{Lowell} {et~al.}(2024){Lowell}, {Jacquemin-Ide}, {Tchekhovskoy}, \&
  {Duncan}}]{Lowell.2024}
{Lowell}, B., {Jacquemin-Ide}, J., {Tchekhovskoy}, A., \& {Duncan}, A. 2024,
  \apj, 960, 82

\bibitem[{{McClintock} {et~al.}(2014){McClintock}, {Narayan}, \&
  {Steiner}}]{2014SSRv..183..295M}
{McClintock}, J.~E., {Narayan}, R., \& {Steiner}, J.~F. 2014, \ssr, 183, 295

\bibitem[{{Miller} {et~al.}(2020){Miller}, {Callister}, \&
  {Farr}}]{2020ApJ...895..128M}
{Miller}, S., {Callister}, T.~A., \& {Farr}, W.~M. 2020, \apj, 895, 128

\bibitem[{{Miller-Jones} {et~al.}(2021){Miller-Jones}, {Bahramian}, {Orosz},
  {Mandel}, {Gou}, {Maccarone}, {Neijssel}, {Zhao}, {Zi{\'o}{\l}kowski},
  {Reid}, {Uttley}, {Zheng}, {Byun}, {Dodson}, {Grinberg}, {Jung}, {Kim},
  {Marcote}, {Markoff}, {Rioja}, {Rushton}, {Russell}, {Sivakoff}, {Tetarenko},
  {Tudose}, \& {Wilms}}]{2021Sci...371.1046M}
{Miller-Jones}, J. C.~A., {Bahramian}, A., {Orosz}, J.~A., {et~al.} 2021,
  Science, 371, 1046

\bibitem[{{Moreno M{\'e}ndez}(2011)}]{2011MNRAS.413..183M}
{Moreno M{\'e}ndez}, E. 2011, \mnras, 413, 183

\bibitem[{{Moreno M{\'e}ndez} \& {Cantiello}(2016)}]{2016NewA...44...58M}
{Moreno M{\'e}ndez}, E. \& {Cantiello}, M. 2016, \na, 44, 58

\bibitem[{{Mosser} {et~al.}(2012){Mosser}, {Goupil}, {Belkacem}, {Marques},
  {Beck}, {Bloemen}, {De Ridder}, {Barban}, {Deheuvels}, {Elsworth}, {Hekker},
  {Kallinger}, {Ouazzani}, {Pinsonneault}, {Samadi}, {Stello}, {Garc{\'\i}a},
  {Klaus}, {Li}, {Mathur}, \& {Morris}}]{2012A&A...548A..10M}
{Mosser}, B., {Goupil}, M.~J., {Belkacem}, K., {et~al.} 2012, \aap, 548, A10

\bibitem[{{Neijssel} {et~al.}(2021){Neijssel}, {Vinciguerra},
  {Vigna-G{\'o}mez}, {Hirai}, {Miller-Jones}, {Bahramian}, {Maccarone}, \&
  {Mandel}}]{2021ApJ...908..118N}
{Neijssel}, C.~J., {Vinciguerra}, S., {Vigna-G{\'o}mez}, A., {et~al.} 2021,
  \apj, 908, 118

\bibitem[{{Nugis} \& {Lamers}(2000)}]{2000A&A...360..227N}
{Nugis}, T. \& {Lamers}, H.~J.~G.~L.~M. 2000, \aap, 360, 227

\bibitem[{{Olejak} \& {Belczynski}(2021)}]{2021ApJ...921L...2O}
{Olejak}, A. \& {Belczynski}, K. 2021, \apjl, 921, L2

\bibitem[{{Orosz} {et~al.}(2011){Orosz}, {McClintock}, {Aufdenberg},
  {Remillard}, {Reid}, {Narayan}, \& {Gou}}]{2011ApJ...742...84O}
{Orosz}, J.~A., {McClintock}, J.~E., {Aufdenberg}, J.~P., {et~al.} 2011, \apj,
  742, 84

\bibitem[{{Orosz} {et~al.}(2007){Orosz}, {McClintock}, {Narayan}, {Bailyn},
  {Hartman}, {Macri}, {Liu}, {Pietsch}, {Remillard}, {Shporer}, \&
  {Mazeh}}]{2007Natur.449..872O}
{Orosz}, J.~A., {McClintock}, J.~E., {Narayan}, R., {et~al.} 2007, \nat, 449,
  872

\bibitem[{{Orosz} {et~al.}(2009){Orosz}, {Steeghs}, {McClintock}, {Torres},
  {Bochkov}, {Gou}, {Narayan}, {Blaschak}, {Levine}, {Remillard}, {Bailyn},
  {Dwyer}, \& {Buxton}}]{2009ApJ...697..573O}
{Orosz}, J.~A., {Steeghs}, D., {McClintock}, J.~E., {et~al.} 2009, \apj, 697,
  573

\bibitem[{{Patton} \& {Sukhbold}(2020)}]{2020MNRAS.499.2803P}
{Patton}, R.~A. \& {Sukhbold}, T. 2020, \mnras, 499, 2803

\bibitem[{{Paxton} {et~al.}(2011){Paxton}, {Bildsten}, {Dotter}, {Herwig},
  {Lesaffre}, \& {Timmes}}]{2011ApJS..192....3P}
{Paxton}, B., {Bildsten}, L., {Dotter}, A., {et~al.} 2011, \apjs, 192, 3

\bibitem[{{Paxton} {et~al.}(2013){Paxton}, {Cantiello}, {Arras}, {Bildsten},
  {Brown}, {Dotter}, {Mankovich}, {Montgomery}, {Stello}, {Timmes}, \&
  {Townsend}}]{2013ApJS..208....4P}
{Paxton}, B., {Cantiello}, M., {Arras}, P., {et~al.} 2013, \apjs, 208, 4

\bibitem[{{Paxton} {et~al.}(2015){Paxton}, {Marchant}, {Schwab}, {Bauer},
  {Bildsten}, {Cantiello}, {Dessart}, {Farmer}, {Hu}, {Langer}, {Townsend},
  {Townsley}, \& {Timmes}}]{2015ApJS..220...15P}
{Paxton}, B., {Marchant}, P., {Schwab}, J., {et~al.} 2015, \apjs, 220, 15

\bibitem[{{Paxton} {et~al.}(2018){Paxton}, {Schwab}, {Bauer}, {Bildsten},
  {Blinnikov}, {Duffell}, {Farmer}, {Goldberg}, {Marchant}, {Sorokina},
  {Thoul}, {Townsend}, \& {Timmes}}]{2018ApJS..234...34P}
{Paxton}, B., {Schwab}, J., {Bauer}, E.~B., {et~al.} 2018, \apjs, 234, 34

\bibitem[{{Paxton} {et~al.}(2019){Paxton}, {Smolec}, {Schwab}, {Gautschy},
  {Bildsten}, {Cantiello}, {Dotter}, {Farmer}, {Goldberg}, {Jermyn}, {Kanbur},
  {Marchant}, {Thoul}, {Townsend}, {Wolf}, {Zhang}, \&
  {Timmes}}]{2019ApJS..243...10P}
{Paxton}, B., {Smolec}, R., {Schwab}, J., {et~al.} 2019, \apjs, 243, 10

\bibitem[{{Pietsch} {et~al.}(2006){Pietsch}, {Haberl}, {Sasaki}, {Gaetz},
  {Plucinsky}, {Ghavamian}, {Long}, \& {Pannuti}}]{2006ApJ...646..420P}
{Pietsch}, W., {Haberl}, F., {Sasaki}, M., {et~al.} 2006, \apj, 646, 420

\bibitem[{{Poutanen} {et~al.}(2007){Poutanen}, {Lipunova}, {Fabrika},
  {Butkevich}, \& {Abolmasov}}]{2007MNRAS.377.1187P}
{Poutanen}, J., {Lipunova}, G., {Fabrika}, S., {Butkevich}, A.~G., \&
  {Abolmasov}, P. 2007, \mnras, 377, 1187

\bibitem[{{Prestwich} {et~al.}(2007){Prestwich}, {Kilgard}, {Crowther},
  {Carpano}, {Pollock}, {Zezas}, {Saar}, {Roberts}, \&
  {Ward}}]{2007ApJ...669L..21P}
{Prestwich}, A.~H., {Kilgard}, R., {Crowther}, P.~A., {et~al.} 2007, \apjl,
  669, L21

\bibitem[{{Qin} {et~al.}(2018){Qin}, {Fragos}, {Meynet}, {Andrews},
  {S{\o}rensen}, \& {Song}}]{2018A&A...616A..28Q}
{Qin}, Y., {Fragos}, T., {Meynet}, G., {et~al.} 2018, \aap, 616, A28

\bibitem[{{Qin} {et~al.}(2019){Qin}, {Marchant}, {Fragos}, {Meynet}, \&
  {Kalogera}}]{2019ApJ...870L..18Q}
{Qin}, Y., {Marchant}, P., {Fragos}, T., {Meynet}, G., \& {Kalogera}, V. 2019,
  \apjl, 870, L18

\bibitem[{{Qin} {et~al.}(2022){Qin}, {Shu}, {Yi}, \&
  {Wang}}]{2022RAA....22c5023Q}
{Qin}, Y., {Shu}, X., {Yi}, S., \& {Wang}, Y.-Z. 2022, Research in Astronomy
  and Astrophysics, 22, 035023

\bibitem[{{Quast} {et~al.}(2019){Quast}, {Langer}, \&
  {Tauris}}]{2019A&A...628A..19Q}
{Quast}, M., {Langer}, N., \& {Tauris}, T.~M. 2019, \aap, 628, A19

\bibitem[{{Ramachandran} {et~al.}(2022){Ramachandran}, {Oskinova}, {Hamann},
  {Sander}, {Todt}, {Pauli}, {Shenar}, {Torrej{\'o}n}, {Postnov}, {Blondin},
  {Bozzo}, {Hainich}, \& {Massa}}]{2022}
{Ramachandran}, V., {Oskinova}, L.~M., {Hamann}, W.~R., {et~al.} 2022, \aap,
  667, A77

\bibitem[{{Reynolds}(2014)}]{2014SSRv..183..277R}
{Reynolds}, C.~S. 2014, \ssr, 183, 277

\bibitem[{{Reynolds}(2021)}]{2021ARA&A..59..117R}
{Reynolds}, C.~S. 2021, \araa, 59, 117

\bibitem[{{Romero-Shaw} {et~al.}(2023){Romero-Shaw}, {Hirai}, {Bahramian},
  {Willcox}, \& {Mandel}}]{2023MNRAS.524..245R}
{Romero-Shaw}, I., {Hirai}, R., {Bahramian}, A., {Willcox}, R., \& {Mandel}, I.
  2023, \mnras, 524, 245

\bibitem[{{Roulet} {et~al.}(2021){Roulet}, {Chia}, {Olsen}, {Dai},
  {Venumadhav}, {Zackay}, \& {Zaldarriaga}}]{2021PhRvD.104h3010R}
{Roulet}, J., {Chia}, H.~S., {Olsen}, S., {et~al.} 2021, \prd, 104, 083010

\bibitem[{{Salvesen} \& {Miller}(2021)}]{2021MNRAS.500.3640S}
{Salvesen}, G. \& {Miller}, J.~M. 2021, \mnras, 500, 3640

\bibitem[{{Sana} {et~al.}(2013){Sana}, {de Koter}, {de Mink}, {Dunstall},
  {Evans}, {H{\'e}nault-Brunet}, {Ma{\'\i}z Apell{\'a}niz},
  {Ram{\'\i}rez-Agudelo}, {Taylor}, {Walborn}, {Clark}, {Crowther}, {Herrero},
  {Gieles}, {Langer}, {Lennon}, \& {Vink}}]{2013A&A...550A.107S}
{Sana}, H., {de Koter}, A., {de Mink}, S.~E., {et~al.} 2013, \aap, 550, A107

\bibitem[{{Schr{\o}der} {et~al.}(2018){Schr{\o}der}, {Batta}, \&
  {Ramirez-Ruiz}}]{2018ApJ...862L...3S}
{Schr{\o}der}, S.~L., {Batta}, A., \& {Ramirez-Ruiz}, E. 2018, \apjl, 862, L3

\bibitem[{{Seifina} \& {Titarchuk}(2010)}]{2010ApJ...722..586S}
{Seifina}, E. \& {Titarchuk}, L. 2010, \apj, 722, 586

\bibitem[{{Sen} {et~al.}(2024){Sen}, {El Mellah}, {Langer}, {Xu}, {Quast}, \&
  {Pauli}}]{2024arXiv240608596S}
{Sen}, K., {El Mellah}, I., {Langer}, N., {et~al.} 2024, arXiv e-prints,
  arXiv:2406.08596

\bibitem[{{Sen} {et~al.}(2021){Sen}, {Xu}, {Langer}, {El Mellah},
  {Sch{\"u}rmann}, \& {Quast}}]{2021A&A...652A.138S}
{Sen}, K., {Xu}, X.~T., {Langer}, N., {et~al.} 2021, \aap, 652, A138

\bibitem[{{Shao} \& {Li}(2022)}]{2022ApJ...930...26S}
{Shao}, Y. \& {Li}, X.-D. 2022, \apj, 930, 26

\bibitem[{{Shimanskii} {et~al.}(2012){Shimanskii}, {Karitskaya}, {Bochkarev},
  {Galazutdinov}, {Lyuty}, \& {Shimanskaya}}]{2012ARep...56..741S}
{Shimanskii}, V.~V., {Karitskaya}, E.~A., {Bochkarev}, N.~G., {et~al.} 2012,
  Astronomy Reports, 56, 741

\bibitem[{{Spruit}(2002)}]{2002A&A...381..923S}
{Spruit}, H.~C. 2002, \aap, 381, 923

\bibitem[{{Sukhbold} {et~al.}(2016){Sukhbold}, {Ertl}, {Woosley}, {Brown}, \&
  {Janka}}]{2016ApJ...821...38S}
{Sukhbold}, T., {Ertl}, T., {Woosley}, S.~E., {Brown}, J.~M., \& {Janka}, H.~T.
  2016, \apj, 821, 38

\bibitem[{{Taylor} \& {Reynolds}(2018)}]{2018ApJ...855..120T}
{Taylor}, C. \& {Reynolds}, C.~S. 2018, \apj, 855, 120

\bibitem[{{Tchekhovskoy} {et~al.}(2012){Tchekhovskoy}, {McKinney}, \&
  {Narayan}}]{Tchekhovskoy.2012}
{Tchekhovskoy}, A., {McKinney}, J.~C., \& {Narayan}, R. 2012, in Journal of
  Physics Conference Series, Vol. 372, Journal of Physics Conference Series,
  012040

\bibitem[{{Thomsen} {et~al.}(2022){Thomsen}, {Kwan}, {Dai}, {Wu}, {Roth}, \&
  {Ramirez-Ruiz}}]{Thomsen22}
{Thomsen}, L.~L., {Kwan}, T.~M., {Dai}, L., {et~al.} 2022, \apjl, 937, L28

\bibitem[{{Vanbeveren} {et~al.}(2020){Vanbeveren}, {Mennekens}, {van den
  Heuvel}, \& {Van Bever}}]{2020A&A...636A..99V}
{Vanbeveren}, D., {Mennekens}, N., {van den Heuvel}, E.~P.~J., \& {Van Bever},
  J. 2020, \aap, 636, A99

\bibitem[{{Vinciguerra} {et~al.}(2020){Vinciguerra}, {Neijssel},
  {Vigna-G{\'o}mez}, {Mandel}, {Podsiadlowski}, {Maccarone}, {Nicholl},
  {Kingdon}, {Perry}, \& {Salemi}}]{2020MNRAS.498.4705V}
{Vinciguerra}, S., {Neijssel}, C.~J., {Vigna-G{\'o}mez}, A., {et~al.} 2020,
  \mnras, 498, 4705

\bibitem[{{Vink} {et~al.}(2001){Vink}, {de Koter}, \&
  {Lamers}}]{2001A&A...369..574V}
{Vink}, J.~S., {de Koter}, A., \& {Lamers}, H.~J.~G.~L.~M. 2001, \aap, 369, 574

\bibitem[{{Winter} {et~al.}(2006){Winter}, {Mushotzky}, \&
  {Reynolds}}]{2006ApJ...649..730W}
{Winter}, L.~M., {Mushotzky}, R.~F., \& {Reynolds}, C.~S. 2006, \apj, 649, 730

\bibitem[{{Xing} {et~al.}(2024){Xing}, {Bavera}, {Fragos}, {Kruckow},
  {Rom{\'a}n-Garza}, {Andrews}, {Dotter}, {Kovlakas}, {Misra}, {Srivastava},
  {Rocha}, {Sun}, \& {Zapartas}}]{2024A&A...683A.144X}
{Xing}, Z., {Bavera}, S.~S., {Fragos}, T., {et~al.} 2024, \aap, 683, A144

\bibitem[{{Zdziarski} {et~al.}(2024{\natexlab{a}}){Zdziarski}, {Banerjee},
  {Chand}, {Dewangan}, {Misra}, {Szanecki}, \&
  {Nied{\'z}wiecki}}]{2024ApJ...962..101Z}
{Zdziarski}, A.~A., {Banerjee}, S., {Chand}, S., {et~al.} 2024{\natexlab{a}},
  \apj, 962, 101

\bibitem[{{Zdziarski} {et~al.}(2024{\natexlab{b}}){Zdziarski}, {Chand},
  {Banerjee}, {Szanecki}, {Janiuk}, {Lubi{\'n}ski}, {Nied{\'z}wiecki},
  {Dewangan}, \& {Misra}}]{2024ApJ...967L...9Z}
{Zdziarski}, A.~A., {Chand}, S., {Banerjee}, S., {et~al.} 2024{\natexlab{b}},
  \apjl, 967, L9

\bibitem[{{Zdziarski} {et~al.}(2013){Zdziarski}, {Mikolajewska}, \&
  {Belczynski}}]{2013MNRAS.429L.104Z}
{Zdziarski}, A.~A., {Mikolajewska}, J., \& {Belczynski}, K. 2013, \mnras, 429,
  L104

\end{thebibliography}
\bibliographystyle{aa}
\nocite{Kwan.prep}



\appendix
\section{Surface and Center Helium Abundance}\label{sec:appendixa}

\begin{figure*}[ht!]
\includegraphics[width=0.96\textwidth]{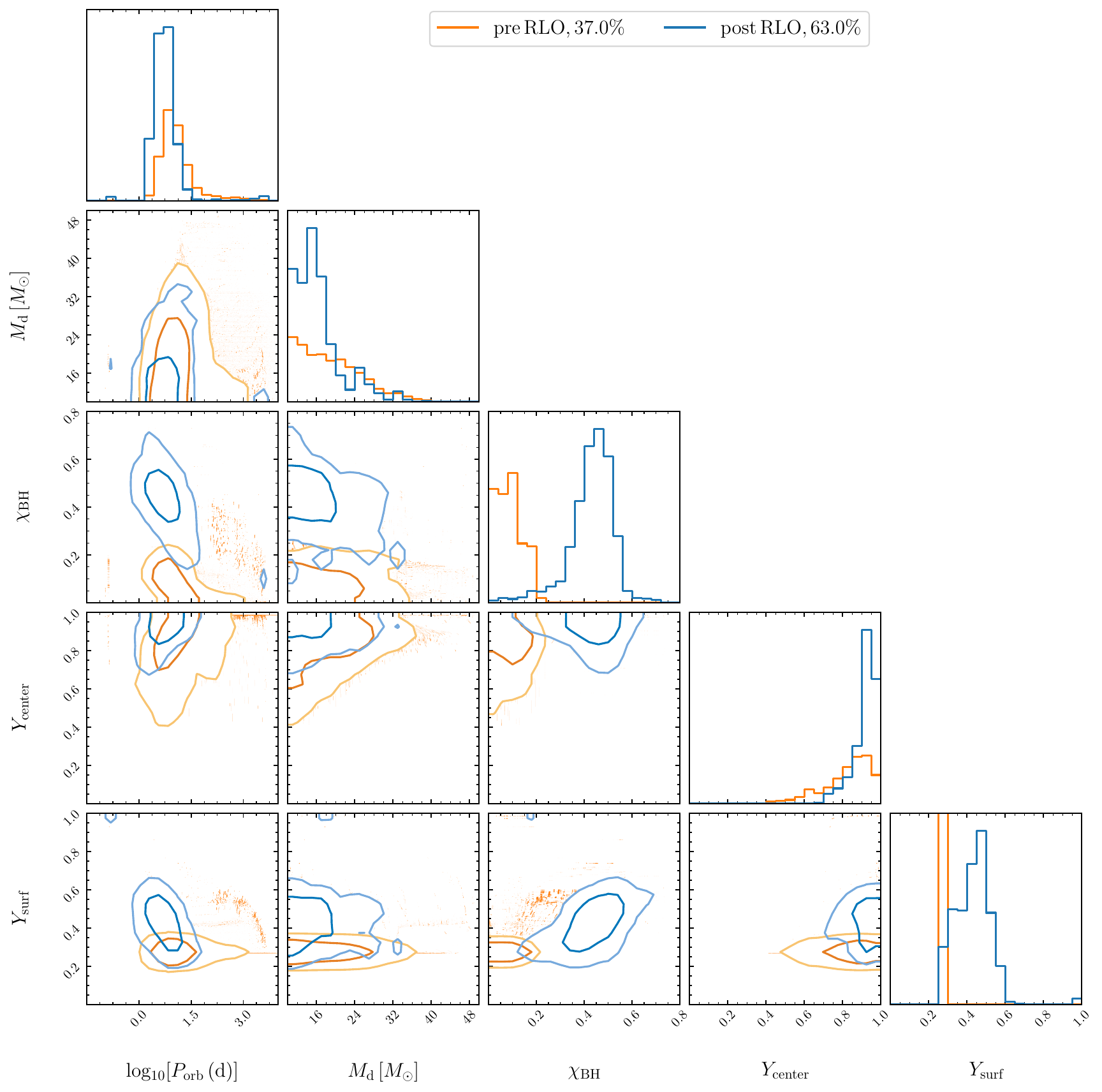}
\caption{Corner plot similar to Figure \ref{fig:super} including the surface and center helium abundance. 
\label{fig:super2}}
\end{figure*}

In this section, we present the population property distributions, in the form of a corner plot, that includes center helium abundance $Y_{\rm{center}}$ and surface helium abundance $Y_{\rm{surf}}$ of the donor star under the moderately super-Eddington accretion scenario. In Figure~\ref{fig:super2}, we show the time-weighted distribution of orbital period, donor star mass, black hole spin, donor star's center helium abundance, and surface helium abundance. For both pre-RLO and post-RLO systems, the majority of the donor stars have $Y_{\rm{center}}$ above $\approx 0.8$, which means that they are approaching the terminal-age MS. As a consequence of RLO, the outer layer of the envelope of the donor stars is stripped, leading to an evident increase in $Y_{\rm{surf}}$. 

\section{Population Property Distributions for Eddington-limited and Fully Conservative BH Accretion} \label{sec:appendixb}

\begin{figure*}[ht!]
\includegraphics[width=0.96\textwidth]{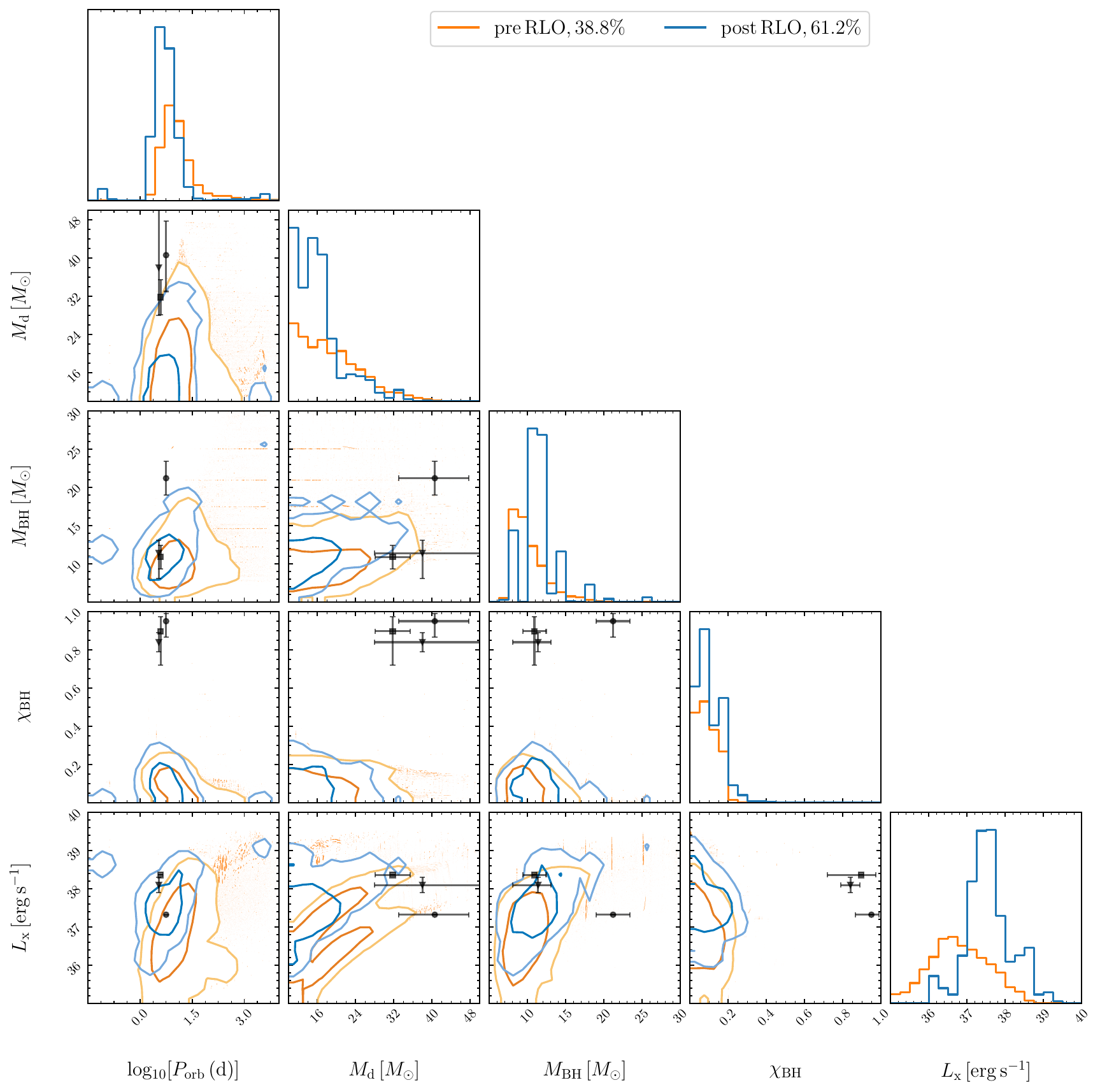}
\caption{Same as Figure \ref{fig:super} but for Eddington-limited BH accretion. 
\label{fig:default}}
\end{figure*}

\begin{figure*}[ht!]
\includegraphics[width=0.96\textwidth]{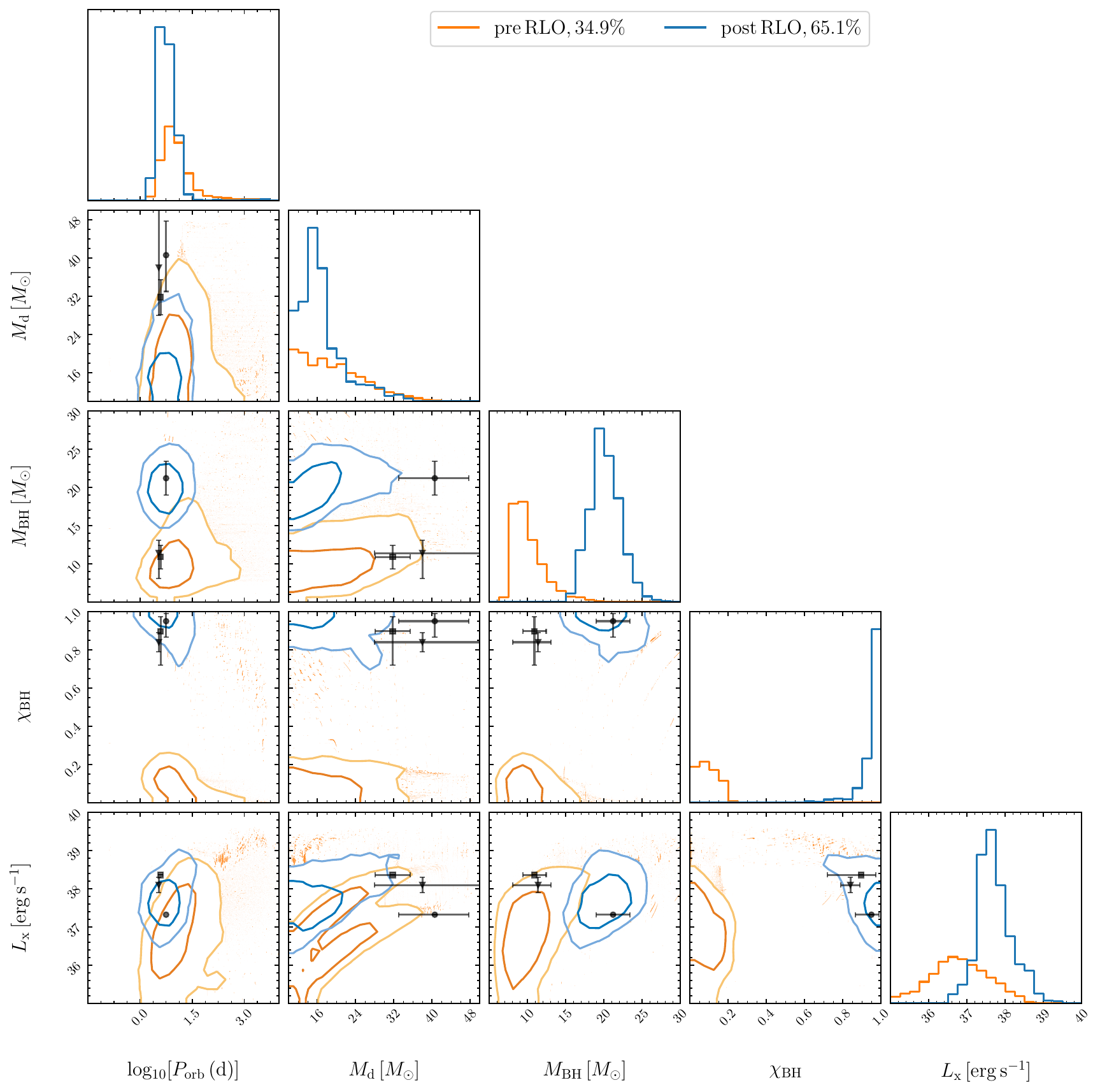}
\caption{Same as Figure \ref{fig:super} but for fully conservative BH accretion. 
\label{fig:full}}
\end{figure*}

Here we display population property distributions, in the form of corner plots, for the population properties of wind-fed BH-HMXBs under the model assumptions of Eddington-limited and fully conservative BH accretion. Figure~\ref{fig:default} shows the same distributions as Figure~\ref{fig:super} but for the Eddington-limited accretion model. The BH spins are $\lesssim 0.3$, with no significant BH spin-up observed to distinguish the populations before and after MT via RLO. Figure~\ref{fig:full} shows the case of fully conservative BH accretion. Bimodal distributions can be seen for BH masses and spins due to accretion. Most of the BHs can be spun up to have spins above $0.8$ after MT via RLO and the binaries remain at this stage longer. The distributions of orbital periods and donor star masses appear similar across different BH accretion models.

\section{Population Property Distributions for X-ray bright sources} \label{sec:appendixc}

\begin{figure*}[ht!]
\includegraphics[width=0.96\textwidth]{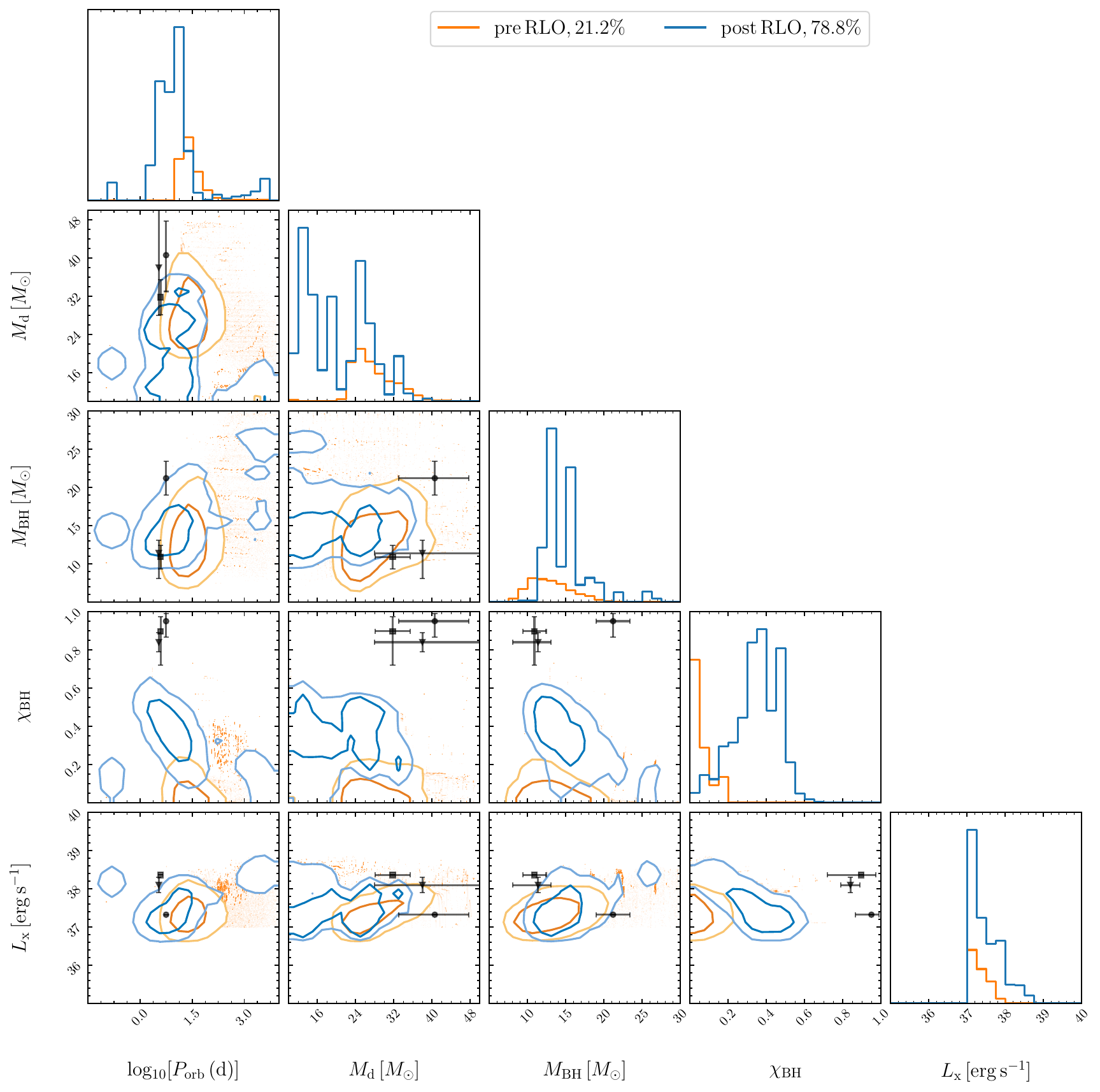}
\caption{Same as Figure \ref{fig:super} but for binaries with X-ray luminosity exceeding $10^{37}\,\rm{erg\,s^{-1}}$, following the application of a low wind accretion efficiency.
\label{fig:x-ray}}
\end{figure*}

When adopting the low-end wind accretion efficiency from \citet{hirai_mandel_2021}, a large fraction of the binaries with donor stars below $20\, M_{\odot}$ exhibit X-ray luminosities below $10^{37}\,\rm{erg\,s^{-1}}$. To illustrate the characteristics of X-ray bright systems, we display a corner plot of the binaries with X-ray luminosity exceeding $10^{37}\,\rm{erg\,s^{-1}}$ in Figure~\ref{fig:x-ray}. About $79\%$ of them are post-RLO systems. The donor stars exhibit a relatively flat distribution from $10$ to $30\,M_{\odot}$, extending up to $\approx 40\,M_{\odot}$. The X-ray bright systems are characterized by a higher proportion of massive donor stars because of their strong winds.

\end{document}